\documentclass[12pt,preprint]{aastex}
\usepackage[]{emulateapj5,onecolfloat}
\usepackage{epsfig}

\begin{document}

\newcommand{\lsim}   {\mathrel{\mathop{\kern 0pt \rlap
  {\raise.2ex\hbox{$<$}}}
  \lower.9ex\hbox{\kern-.190em $\sim$}}}
\newcommand{\gsim}   {\mathrel{\mathop{\kern 0pt \rlap
  {\raise.2ex\hbox{$>$}}}
\lower.9ex\hbox{\kern-.190em $\sim$}}}
\def\be{\begin{equation}}
\def\ee{\end{equation}}
\def\ba{\begin{eqnarray}}
\def\ea{\end{eqnarray}}
\def\eps{{\varepsilon}}
\def\ap{\approx}
\def\bb{\leftarrow}

\title{
{\bf Diffusive propagation of UHECR and the propagation theorem}}

\author{R. Aloisio and V. Berezinsky}
\affil{INFN - Laboratori Nazionali del Gran Sasso, I--67010
Assergi (AQ), Italy}

\email{aloisio@lngs.infn.it, berezinsky@lngs.infn.it}

\begin{abstract}
We present a detailed analytical study of the propagation of ultra
high energy (UHE) particles in extragalactic magnetic fields.
The crucial parameter which affects the diffuse spectrum is the
separation between sources. In the case of a uniform distribution
of sources with a separation between them much smaller than all
characteristic propagation lengths, the diffuse spectrum of UHE
particles has a {\em universal} form, independent of the mode of
propagation. This statement has a status of theorem. The proof is
obtained using the particle number conservation during
propagation, and also using the kinetic equation for the
propagation of UHE particles. This theorem can be also proved with
the help of the diffusion equation. In particular, it is shown
numerically, how the diffuse fluxes converge to this universal
spectrum, when the separation between sources diminishes. We study
also the analytic solution of the diffusion equation in weak and
strong magnetic fields with energy losses taken into account. In
the case of strong magnetic fields and for a separation between
sources large enough, the GZK cutoff can practically disappear, as it
has been found early in numerical simulations. In practice,
however, the source luminosities required are too large for this 
possibility.
\end{abstract}
\section{Introduction}
\label{introduction}

The propagation of UHECR protons and nuclei with $E\gsim 1\times 10^{19}$~eV
in the large-scale intergalactic magnetic field (IMF) remains an open
problem, mainly because the knowledge of the IMF is still very poor. The
possibilities vary between rectilinear propagation in a weak field and
diffusive propagation in a strong magnetic field. The experimental data on
IMF and the models of origin of these fields do not
allow at present to choose even between the two extreme propagation regimes
mentioned above.

Most reliable observations of the intergalactic magnetic field are
based on the Faraday rotation of the polarized radio emission (for
the reviews see Kromberg (1994), Vall\'e (1997), Carilli and
Taylor (2002)) . The upper limit on the Faraday rotation measure
(RM) in the extragalactic magnetic field, obtained from the
observations of distant quasars, gives an upper limit of ${\rm RM} <
5~{\rm rad/m}^2$. It implies an upper limit on the extragalactic
magnetic field on each assumed scale of coherence length (Kromberg (1994),
Vall\`e (1997), Ryu et al. (1998)). For example, according to 
Blasi et al. (1999a) for an inhomogeneous universe
$B_{l_c} < 4$~nG on a scale of coherence $l_c= 50$~Mpc.

According to observations of the Faraday rotations the
extragalactic magnetic field is strongest, or order of $1~\mu$G,
in clusters of galaxies and radiolobes of radiogalaxies
(Vall\'e (1997), Kromberg (1994), Carilli and Taylor (2002)). The
largest scale in both structures reaches $l_c \sim 1$~ Mpc. Most
probably various structures of the universe differ dramatically by
magnetic fields, with very weak field in voids and much stronger
in the filaments (Ryu et al. (1998)). Superclusters seem to be too
young for the regular magnetic field to be formed in these structures on a
large scale $l_c \sim 10$~Mpc.

In case of hierarchical magnetic field structures in the universe,
UHE protons with $E> 4\times 10^{19}$~eV can propagate in a
quasi-rectilinear regime. Scattering of UHE protons occurs mostly
in galaxy clusters, radiolobes and filaments. Deflections of UHE
protons can be large for some directions and small for the others.
The universe looks like a leaky, worm-holed box, and correlation
with the sources can be observable (see Tinyakov and Tkachev (2001)),
where correlations of UHECR with BLLacs are found). Such a
picture has been suggested by Berezinsky et al. (2002b).

A promising theoretical tool to predict the IMF in large scale
structures is given by magneto-hydrodynamic (MHD) simulations.
The main uncertainty in these simulations is related to the
assumptions concerning the seed magnetic field.

The MHD simulations of Sigl et al. (2004) and Sigl et al. (2003)
favor the hierarchical structure with strong magnetic fields.
Assuming an inhomogeneous seed magnetic field generated by cosmic
shocks through the Biermann battery mechanism, the authors obtain
a $\sim 100$~nG magnetic field in filaments and $\sim 1$~nG in
voids. In some cases they consider IMF up to a few micro Gauss as
allowed. In these simulations UHECR are characterized by large
deflection angles, of the order of $20^{\circ}$, at energies
up to $E\sim 10^{20}$~eV (Sigl et al. (2003), Sigl et al. (2004)). Thus,
the scenario that emerges in these simulations seems to exclude
the UHECR astronomy. These simulations have some ambiguity related
to the choice of magnetic field at the position of the observer
(Sigl et al. (2003), Sigl et al. (2004)). The authors consider two
cases: a strong local magnetic field $B\sim 100$~nG and a weak
field $B \ll 100$~nG. The different assumptions about the local
magnetic field strongly affects the conclusions about UHECR
spectrum and anisotropy.

The essential step forward in MHD simulations has been made
recently. In Dolag et al. (2003) the Local Universe is simulated
with the observed density and velocity field. This
eliminates the ambiguity for the local magnetic field, that 
is found to be weak. The seed magnetic field, used in
this simulation, is normalized by the observed magnetic field in
rich clusters of galaxies. The results of these constrained
simulations indicate the weak magnetic fields in the universe of
the order of $0.1$~nG in typical filaments and of $0.01$~nG in
voids. The strong large-scale magnetic field, $B\sim 10^{3}$~nG,
exists in clusters of galaxies, which, however, occupy
insignificant volume of the universe. The picture that emerges
from simulations of Dolag at el. (2003) favors a hierarchical
magnetic field structure characterized by weak magnetic fields.
UHE protons with $E> 4\times 10^{19}$~eV can propagate in a
quasi-rectilinear regime, with the expected deflection angles being very
small $\le 1^{\circ}$. However, until direct observational
evidences for this picture becomes available, an alternative case
of propagation in strong magnetic fields, with diffusion as
an extreme possibility, can be hardly excluded.

This case has been studied in Sigl et al. (1999), Lemoine et al.
(1999), Stanev (2000), Harari et al. (2002), Yoshiguchi et al.
(2003), Deligny et al. (2003). An interesting features found in
these calculations are small-angle clustering of UHE particles as 
observed by Hayashida et al. (1996), Hayashida et al. (1999), 
Uchiori et al. (2000), Glushkov and Pravdin (2001), 
and absence of the GZK cutoff in the 
diffusive propagation, when the magnetic field is very strong. Many 
aspects of diffusion of UHECR have been studied in numerical simulation 
by Casse et al. (2002).

We shall illustrate UHECR propagation in strong magnetic fields by
the calculations by Yoshiguchi et al. (2003). The authors performed MC
simulations for propagation in random magnetic field with the
Kolmogorov spectrum of turbulent energy density $w_k \propto k^n$,
with $n=-5/3$. The basic (largest) coherent scale is chosen as
$l_c$ with the  mean field $B_0$. Numerically these parameters
vary in the range (1 - 40)~Mpc for $l_c$, and (1 - 100)~nG - for
$B_0$. The sources are taken as galaxies from Optical Redshift
Survey catalog with absolute magnitude $M_B$ brighter than some
critical value $M_c$. The calculated quantities are the energy
spectrum, anisotropy and small-angle clustering.  The observed
small-angle clustering and absence of the GZK cutoff in the AGASA
observations can be reproduced in the case of strong magnetic field 
$B \geq 10$~nG.

Diffusive propagation of extragalactic UHECR has been studied
already in earlier work. The stationary diffusion from Virgo
cluster was considered by Wdowczyk and Wolfendale (1979),
Giller et al. (1980) and non-stationary
diffusion from a nearby source was studied by Berezinsky et al.
(1990a), Blasi and Olinto (1999b) using the the Syrovatsky solution 
(Syrovatskii (1959)) of the diffusion equation. In this case the GZK 
cutoff can be absent. A very similar problem was considered again more 
recently by Isola et al. (2002)).

In this paper we shall study how propagation influences the
diffuse energy spectrum of UHECR. We shall prove the theorem that
if distance between sources  is much smaller than all propagation
lengths, the spectrum has the same universal form independent of the mode
of propagation. For diffusion in magnetic fields we shall demonstrate
how the spectra converge to the universal one when the separation
between sources diminishes. Finally, we shall obtain, with the help of
the Syrovatsky solution, the spectra for strong magnetic field. In this
case with large enough separation between sources, the GZK cutoff becomes 
weak or absent, and we will discuss the physical explanation of this 
phenomenon.

\section{Propagation Theorem}
\label{theorem}

Let us consider a case when identical  UHECR sources are distributed
uniformly \footnote{In this paper we shall distinguish between {\em uniform}
and {\em homogeneous} distribution of the sources: under the latter we
assume a continuous and distance-independent distribution.}
in the space, with $d$ being the separation between sources. We will
demonstrate that if $d$ is less than all other characteristic lengths of
propagation, such as diffusion length $l_d(E)$ and energy attenuation length
$l_{att}(E)$ given by
\be
\label{att}
l_{att}=cE/(dE/dt) ,
\ee
then the diffuse energy spectrum has a universal (standard) form
{\em independent of the mode of particle propagation}. In particular, 
under the conditions specified above the magnetic field, both weak and strong,
does not affect the shape of the energy spectrum.

Explicitly, this theorem can be formulated as follows:

{\em For a uniform distribution of identical sources with
separation much less than the characteristic propagation lengths,
the diffuse spectrum of UHECR has a universal (standard) form,
independent of the mode of propagation}.

First, we shall consider the proof based on the conservation of the
number of particles
(e.g. protons or nuclei) during the propagation.
Let $t$ be the age of the universe with the present age taken as $t_0$.
The number of particles per unit volume of the present universe is equal
to the number of particles injected into this volume during all history of
the universe, independent of the mode of propagation. The homogeneity of
particles needed for this statement is provided by
almost homogeneous distribution of sources.
Thus, the comoving space density of particles
$n_p(E)$ from uniformly distributed sources with an age-dependent
comoving density $n_s(t)$ and age-dependent generation rate by a source,
$Q(E,t)$, is given by
\be
\label{n_p}
n_p(E) dE = \int_0^{t_0} dt Q(E_g,t) n_s(t) dE_g,
\ee
where $E_g(E,t)$ is the required  generation energy at age $t$,
if the observed energy is $E$. Eq.~(\ref{n_p}) does not depend on the 
way particles propagate.

The homogeneous distribution of particles in presence of inhomogeneous
magnetic fields follows from the Liouville theorem and can be explained in
the following way. Suppose we have an observer in the space with
a magnetic field. The diffuse flux in any  direction is given by
the integral $\int n_s(l)dl$ over the trajectory of a particle (or antiparticle
emitted from the observation point). If $n_s$ is almost homogeneous,
the integral depends on the time of propagation and does not depend on
the strength and inhomogeneity of magnetic field.

To find the explicit form of the universal spectrum $n_p(E)$, one needs
some additional assumptions. Let us consider the protons as primaries with
continuous energy losses due to interaction with the CMB radiation
\be
dE/dt= -b(E,t).
\label{E_loss}
\ee
Here and everywhere below $b(E,t)>0$. We shall use the connection between
the redshift $z$ and cosmological time $t$ according to the standard cosmology
\be
\label{dt}
dt = \frac{dz}{H_0(1+z) \sqrt{(1+z)^3\Omega_m + \Omega_{\Lambda}}},
\ee
with $H_0, \Omega_m$ and $\Omega_{\Lambda}$ being the Hubble constant, relative
cosmological density of matter and relative density of vacuum energy, 
respectively.

The generation rate is assumed to be the same for all sources and is
taken in the form
\be
\label{Q}
Q(E,t) = L_p (1+z)^{\alpha} K(\gamma_g) q_{gen}(E_g)
\ee
where $L_p$ is the CR luminosity of the source with $(1+z)^{\alpha}$
describing the possible cosmological evolution of luminosity. The
normalization factor $K(\gamma_g)$ is $\gamma_g - 2$ if $\gamma_g>2$
and $1/\ln (E_{\rm max}/E_0)$ if $\gamma_g=2$, with $E_0$ and $E_{\rm max}$
the minimum and maximum generation energies, respectively. The comoving
density $n_s(t)$ of the sources can also contain the evolutionary factor
$(1+z)^{\beta}$, with $\alpha+\beta=m$. Here and everywhere else we assume
$E_0=1$~GeV, all energies $E$ are measured in GeV and $L_p$ in GeV/s.

Then from Eq. (\ref{n_p}) we obtain the normalized universal
spectrum as:
\be 
\label{u-spectrum}
J_p(E)=\frac{c}{4\pi} {\cal
L}_0 K(\gamma_g)\int_0^{z_{\rm max}}dz \left (\frac{dt}{dz}\right
) (1+z)^{m} q_{gen}(E_g)\frac{dE_g}{dE}
\ee
where $J_p$ is diffuse
flux, ${\cal L}_0=L_p n_s$ is the emissivity at $z=0$, $dt/dz$ is
given by Eq.~(\ref{dt}), $E_g$ is calculated as $E_g=E_g(E,z)$ and
$dE_g/dE$ is given by (Berezinsky and Grigorieva (1998),
Berezinsky et al. (2002a))
\be 
\label{dE_g/dE} 
\frac{d E_g(z_g)}{dE} = (1+z_g) exp \left\{\frac{1}{H_0} \int_0^{z_g} dz
\frac{(1+z)^2}{\sqrt{\Omega_m(1+z)^3+\Omega_{\Lambda}}} \left (
\frac{db_0(E')}{dE'}\right )_{E'=(1+z)E_g(z)} \right \}, 
\ee
where $b_0(E)$ is the proton energy loss at $z=0$.

The generation spectrum $q_{gen}(E_g)$ is not known apriori. Three
kinds of spectrum might be considered: (i) the power law spectrum
$q_{gen}(E_g)=E_g^{-\gamma_g}$ with $\gamma_g>2$ and the
normalization factor $K(\gamma_g)=\gamma_g - 2$, (ii) the
power-law spectrum $q_{gen}(E_g)=1/E_g^2$ with
$K(\gamma_g)=1/\ln(E_{max}/E_0)$ and (iii) the complex spectrum
\begin{equation}
q_{\rm gen}(E_g)=\left\{ \begin{array}{ll}
1/E_g^2                      ~ &{\rm at}~~ E_g \leq E_c\\
E_c^{-2}(E_g/E_c)^{-\gamma_g}~ &{\rm at}~~ E_g \geq E_c
\end{array}
\right.
\label{complex}
\end{equation}
with
$$K(\gamma_g)=\frac{1}{\ln\frac{E_{max}}{E_c} +\frac{1}{\gamma_g-2}}.$$
In numerical calculations with the complex spectrum
(\ref{complex}) we use $\gamma_g=2.7$, which gives the best fit to
observational data (Berezinsky et al. (2003)).

\begin{figure}[t!]
\begin{center}
\begin{tabular}{ll}
\includegraphics[width=0.45\textwidth]{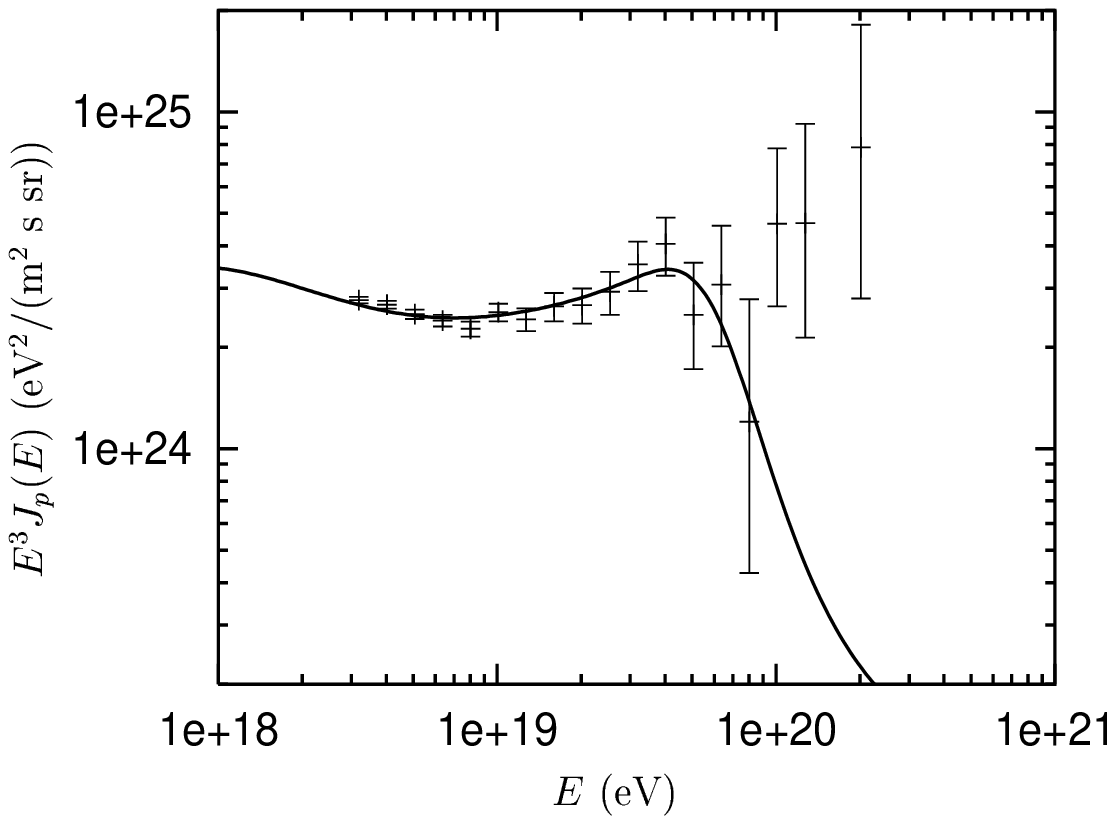}
&
\includegraphics[width=0.45\textwidth]{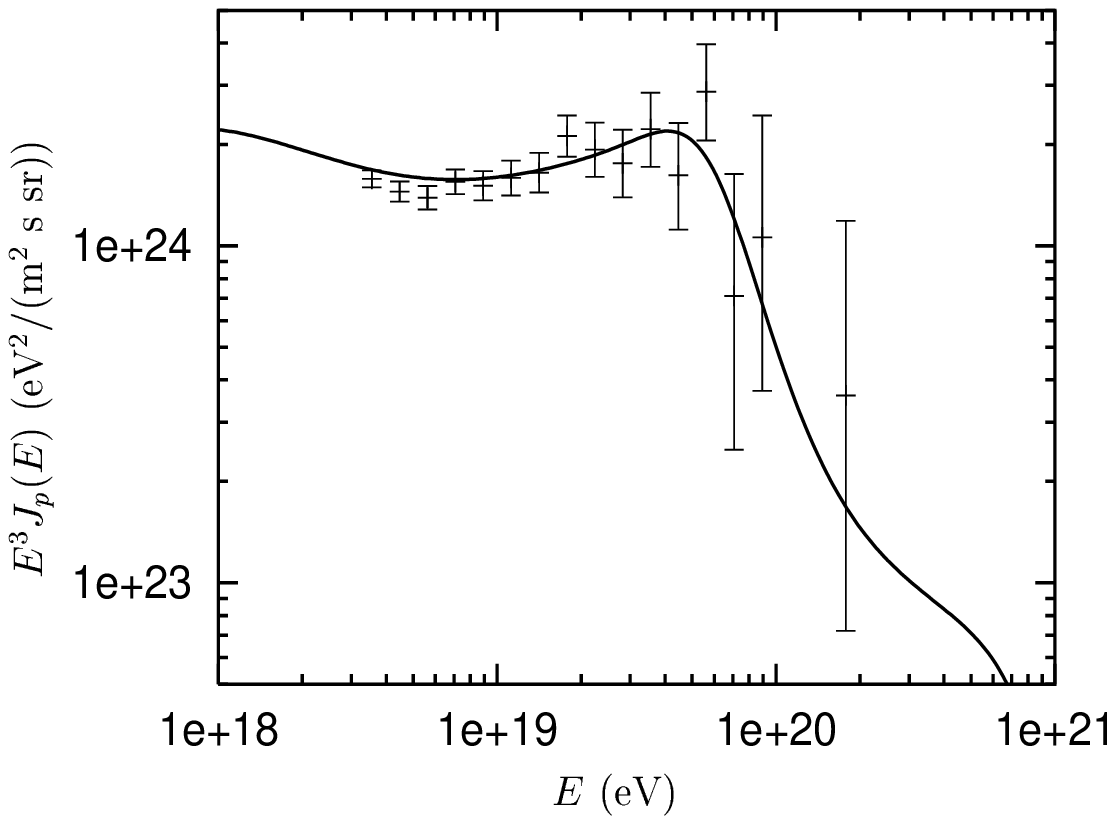}
\\
\end{tabular}
\caption{ Universal spectrum in comparison with the AGASA (left
panel) and HiRes data (right panel).} \label{fig1}
\end{center}
\end{figure}

The universal spectrum (\ref{u-spectrum}) with $q_{\rm gen}$ from
Eq.~(\ref{complex}) and with $m=0$ is shown in Fig. \ref{fig1} in
comparison with experimental data of AGASA and HiRes arrays. From
these figures one can see the good agreement of the universal
spectrum with experimental data of both detectors up to $8\times
10^{19}$~eV. The required emissivities ${\cal L}_0$ are: ${\cal
L}_0\simeq 1.8 \times 10^{46}$~erg/Mpc$^3$yr and ${\cal L}_0\simeq
8.9 \times 10^{45}$~erg/Mpc$^3$yr for the AGASA and HiRes data,
respectively. The excess of AGASA events at $E> 1\times
10^{20}$~eV needs another component of UHECR.

Under the condition $d<<l_{\rm prop}$ the energy spectrum (\ref{u-spectrum})
is the same (universal) for rectilinear propagation and propagation in weak
and strong magnetic fields. If $l_{\rm prop}\le d$ (or $l_{\rm prop}<<d$)
the propagation theorem is not valid any more. The distortion of the
universal spectrum occurs at $E\ge E_{GZK}$ due to several effects:
inhomogeneity of distribution of UHECR sources (local enhancement or local
absence of the sources), fluctuation in the distribution of sources,
fluctuations of the photopion energy losses over the distance $d$ and 
due to the small diffusion length $l_d<d$ in case of strong magnetic fields.
\section{Kinetic equation and universal spectrum}
\label{kinetic}

We have obtained the universal spectrum above from the very general condition
of particles conservation during propagation.
In this Section we shall demonstrate that the universal spectrum follows
also from the kinetic equation.

Let us consider the kinetic equation
for a {\em homogeneous} distribution of the sources (see footnote $^1$).
Since in this case
the particle distribution is also homogeneous, the diffusion term in
the kinetic equation is absent, and it has the form:
\begin{equation}
\label{kin-eq1}
\frac{\partial n_p(E,t)}{\partial t} =
\frac{\partial}{\partial E}\left [b(E,t)n_p(E,t)\right ]+ Q_g(E,t),
\end{equation}
with $Q_g= n_s Q$.
Let us start with a simple stationary equation when $Q_g$, $n_p$ and $b$
do not depend on time:
\begin{equation}
-\frac{\partial}{\partial E}\left [ b(E)n_p(E)\right ] =Q_g(E)
\label{kin-eq2}
\end{equation}
The solution of Eq. (\ref{kin-eq2}) is
\begin{equation}
\label{solut1}
n_p(E)=\frac{1}{b(E)}\int_{E}^{E_{\rm max}}dE_g Q_g(E_g)
=\int dt \frac{b(E_g)}{b(E)}Q_g(E_g),
\end{equation}
where $E_g(E,t)$ is the generation energy at time $t$ and we used
$dE_g=-b(E_g)dt$. Using $dE=-b(E)dt$ for the same interval $dt$ we
obtain $b(E_g)/b(E)=dE_g/dE$ and
\be
n_p(E)=\int dt Q_g(E_g(t))\frac {dE_g}{dE},
\label{solut2}
\ee
in agreement with Eqs. (\ref{n_p}) and (\ref{u-spectrum}).

Let us now come back to Eq. (\ref{kin-eq1}) with time dependent 
quantities $Q_g, n_p$ and $b$, where $b(t)$ includes also adiabatic energy
losses due to redshift. Introducing the new quantities:
$\tilde{b}(E,t)=b(E)/H(t)$,~ $\tilde{Q}_g(E,t)=Q_g(E,t)/H(t)$ and
$\tau=(1+z_{max})/(1+z)$, where $z_{max}$ is the maximal redshift in
the evolution of sources, we obtain the equation
\be
\label{kin-eq3}
\frac{\partial n_p(E,\tau)}{\partial \tau}=
- \frac{\partial \tilde{b}(E,\tau)}{\partial E}n_p(E,\tau)+
\tilde{Q}_p(E,\tau).
\ee
Eq. (\ref{kin-eq3}) may be solved with the help of an integration factor
\be
\label{int-factor}
\mu(\tau) =  \exp \left( - \int_0^\tau d\tau^\prime \frac{\partial
\tilde{b}}{\partial E}(\tau^\prime) \right)
\ee
Assuming $n_p(E,z_m)=0$, we obtain
\be
n_p(E,\tau) = \frac{1}{\mu(\tau)} \left[
\int_0^\tau d\tau^\prime \mu(\tau^\prime) \tilde{Q}_g(E,\tau^\prime) \right].
\ee
At this stage the energy losses should be separated into those due to
redshift and those due to the interaction with the CMB radiation
\be
b(E',z)= E' H_0 \sqrt{\Omega_m(1+z)^3+\Omega_{\Lambda}}+ (1+z)^2 b_0^{\rm CMB}
\left ( E'(1+z)\right )
\label{b}
\ee
where $E'$ is an arbitrary energy at epoch $z$.

Coming back to the variable $z$ in Eq.~(\ref{kin-eq3}) and inserting
there $\partial b(E,z)/\partial E$ from Eq.~(\ref{b}) we obtain after
integration
$$
n_p(E) = \frac {1}{H_0}\int_0^{z_{max}}
\frac{dz}{\sqrt{\Omega_m(1+z)^3+\Omega_{\Lambda}}}Q_g(E_g,z)\times
$$
\begin{equation}
\times \exp \left[ \frac{1}{H_0} \int_0^z
\frac{dz^{\prime}(1+z^{\prime})^2}
{\sqrt{\Omega_m (1 + z^{\prime})^3 + \Omega_{\Lambda}}}
\left( \frac{\partial b_0(E^{\prime},z^{\prime})}{\partial E^{\prime}}
\right)_{E^{\prime}=(1+z)E_g(z^{\prime})}\right] ,
\label{u-spectrum1}
\end{equation}
which, after introducing the explicit expression for $Q_g$ from Eq.~(\ref{Q}),
coincides exactly with Eqs.~(\ref{u-spectrum}) and (\ref{dE_g/dE}).

\section{Diffusion equation and the Syrovastsky solution}
\label{syrovatsky}

In the previous section we have considered the case of a homogeneous
distribution of sources, and proved that
the universal spectrum is a solution of the kinetic equation (\ref{kin-eq1}).
We shall now assume a non-homogeneous distribution of the sources.
Then one should add the propagation term in Eq.~(\ref{kin-eq1}).
We assume the diffusive propagation in large-scale magnetic fields.
For a single source at point $\vec{r}_g$ the diffusion equation has the
form:
\be
\label{diff-eq}
\frac{\partial n_p(E,r)}{\partial t} - {\rm div} \left [ D(E,r,t)\nabla
n_p(E,r)\right]
-\frac{\partial}{\partial E}\left [ b(E)n_p(E,r)\right ] =
Q(E,t)\delta(\vec{r}-\vec{r}_g)
\ee
where $D(E,r,t)$ is the diffusion coefficient that depends on the magnetic
field structure. We refer the reader to the Section \ref{diff-par} for a
detailed discussion of the relation between the diffusion coefficient and
the magnetic field.

In the case when $D$, $b$ and $Q$ depend only on energy, the exact
analytic solution of Eq.~(\ref{diff-eq}), found by
Syrovatskii (1959), is
\begin{equation}
\label{syr-sol} 
n_p(E,r)= \frac{1}{b(E)} \int_{E}^{\infty} dE_g Q(E_g) \frac{exp\left
[-\frac{r^2}{4\lambda(E,E_g)} \right ]} {\left (
4\pi\lambda(E,E_g)\right )^{3/2}}, 
\end{equation}
where 
\be 
\label{lambda}
\lambda(E,E_g) = \int_{E}^{E_g} d\epsilon
\frac{D(\epsilon)}{b(\epsilon)}
\ee
is the Syrovatsky variable
which has the meaning of the squared distance traversed by a
proton in the observer direction, while its energy diminishes from
$E_g$ to $E$.

According to the propagation theorem, integrating the Syrovatsky
solution over homogeneously distributed sources with density
$n_s$,
\begin{equation}
\label{diffuse}
n_p(E)=\int_0^{\infty} dr 4\pi r^2
\frac{n_s}{b(E)}\int_{E}^{\infty} dE_g Q(E_g) \frac{exp\left
[-\frac{r^2}{4\lambda(E,E_g)} \right ]} {\left (
4\pi\lambda(E,E_g)\right )^{3/2}}~,
\end{equation}
must result in the universal spectrum. One easily sees this by
changing the order of integration in Eq. (\ref{diffuse}) and using
$$ \int_0^{\infty}dr \frac{4\pi r^2}{\left (4\pi \lambda \right )^{3/2}}
exp\left [-\frac{r^2}{4\lambda} \right ] = 1~, $$

which gives the diffuse space density of protons
\begin{equation}
\label{diffuse1}
n_p(E)=\frac{n_s}{b(E)} \int_E^{\infty} dE_g Q(E_g)
\end{equation}
in agreement with Eq. (\ref{solut1}), where $Q_g(E)=n_s Q(E)$

Coming back to the non-homogeneous case, let us consider a lattice
with total size $a$ and UHECR sources located in the lattice
vertexes separated by a distance $d$. Using this model for the
source distribution, we shall demonstrate the consistency of the
Syrovatsky solution with the propagation theorem. We will show
that when the separation between sources, $d$, tends to a small
value $d_{lim}$, the Syrovatsky solution gives the universal
spectrum.

Using the lattice distributed sources and the Syrovatsky
solution, one obtains the diffuse flux as 
\be 
\label{Sy-diffuse}
J_p(E)=\frac{c}{4\pi}\frac{1}{b(E)} \sum_i \int_{E}^{E_{max}} dE_g
Q(E_g) \frac{exp\left [-\frac{r_i^2}{4\lambda(E,E_g)} \right ]}
{\left ( 4\pi\lambda(E,E_g)\right )^{3/2}} 
\ee 
where $\lambda(E,E_g)$ is given by Eq. (\ref{lambda}) and the summation goes
over all sources in the lattice vertexes. The maximum energy in
Eq.(\ref{Sy-diffuse}) is given by the smaller of two quantities:
the maximum acceleration energy $E_g^{\rm max}$, and the
generation energy $E_g(E,ct_0)$ of a proton with present energy $E$
propagating during a time $t_0$. For the energies $E$ at interest, 
the former is always smaller that the latter, and we use as $E_{\rm max}$ 
the maximum acceleration energy.

We have to specify now the diffusion coefficient $D(E)$, which
determines $\lambda(E,E_g)$ in Eq. (\ref{Sy-diffuse}). Putting off
the detailed discussion until the Section \ref{diff-par}, we will
give here a short description of the diffusion coefficients used
in this work.

We assume diffusion in a random magnetic field with the mean value
$B_0$ on the maximum coherent length $l_c$. This assumption determines
the diffusion coefficient $D(E)$ at the highest energies when the proton Larmor
radius, $r_L(E) \gg l_c$:
\be
D(E) = \frac{1}{3} \frac{c r^2_L(E)}{l_c}
\label{D_HE}
\ee
At ``low'' energies, when $r_L(E) \lsim l_c$ we shall consider
three cases:

(i) energy-independent diffusion coefficient
\be
\label{D_const}
D = \frac{1}{3} c l_c,
\ee

(ii) the Bohm diffusion coefficient, which provides the lowest value
of $D$
\be
\label{D_B}
D_B(E) = \frac{1}{3} c r_L(E),
\ee

(iii) the Kolmogorov diffusion coefficient
\be
\label{D_K}
D_K(E)= \frac{1}{3}cl_c\left (\frac{r_L(E)}{l_c}\right )^{1/3}.
\ee

In all three cases we normalized the diffusion coefficients by
$(1/3)cl_c$ at $r_L=l_c$ (see Section \ref{diff-par}).
The characteristic energy $E_c$ of the transition between the 
high energy and low energy regimes is determined by the condition
$r_L(E)=l_c$ and is
\be
\label{E_c}
E_c = 0.93\times 10^{18} \left (\frac{B_0}{1~{\rm nG}} \right)
\left (\frac{l_c}{Mpc} \right)~ {\rm eV.}
\ee

One can describe the low-energy and high-energy diffusion regimes
with the help of an interpolation formula for the diffusion length:
\be
\label{l_d}
l_d(E)=\Lambda_d + \frac{r_L^2(E)}{l_c}
\ee
with  $\Lambda_d=l_c$ for the regime with $D=const$, 
$\Lambda_d=r_L(E)$ for the Bohm diffusion and 
$\Lambda_d=l_c (r_L/l_c)^{1/3}$ for the Kolmogorov regime.
For the later use, we shall formalize the description of these 
three regimes using
\begin{equation}
\Lambda_d(E)=l_c(r_L/l_c)^{\alpha},
\label{alpha}
\end{equation}
with $\alpha$ equal to 0, 1 and 1/3 for the $D=const$, Bohm and 
Kolmogorov regimes.

For completeness we shall give also the numerical expression for
the Larmor radius:
\be
\label{r_L}
r_L(E)= 1.08 \times 10^2 \frac{E}{1\times 10^{20}~{\rm eV}}
\frac{1~{\rm nG}}{B}~ {\rm Mpc}.
\ee

One can see, therefore, the existence of two different propagation 
lengths that should be compared with the distance $d$ between sources. 
These two lengths are $l_{att}$ given by Eq.~(\ref{att}) and $l_d$ given by
Eq.~(\ref{l_d}). The former is large even at the highest energies
($l_{att}\sim 25$~Mpc at $E\sim 10^{21}$~eV) while the latter can be small
enough at the energies of interest $10^{19} \le E \le 10^{21}$~eV. For a
representative case $B_0=100$~nG and $l_c=1$~Mpc we have
$l_d(E_c)=l_c=1$~Mpc at $E_c \approx 1\times 10^{20}$~eV.

\begin{figure}[t!]
\begin{center}
\begin{tabular}{ll}
\includegraphics[width=0.45\textwidth]{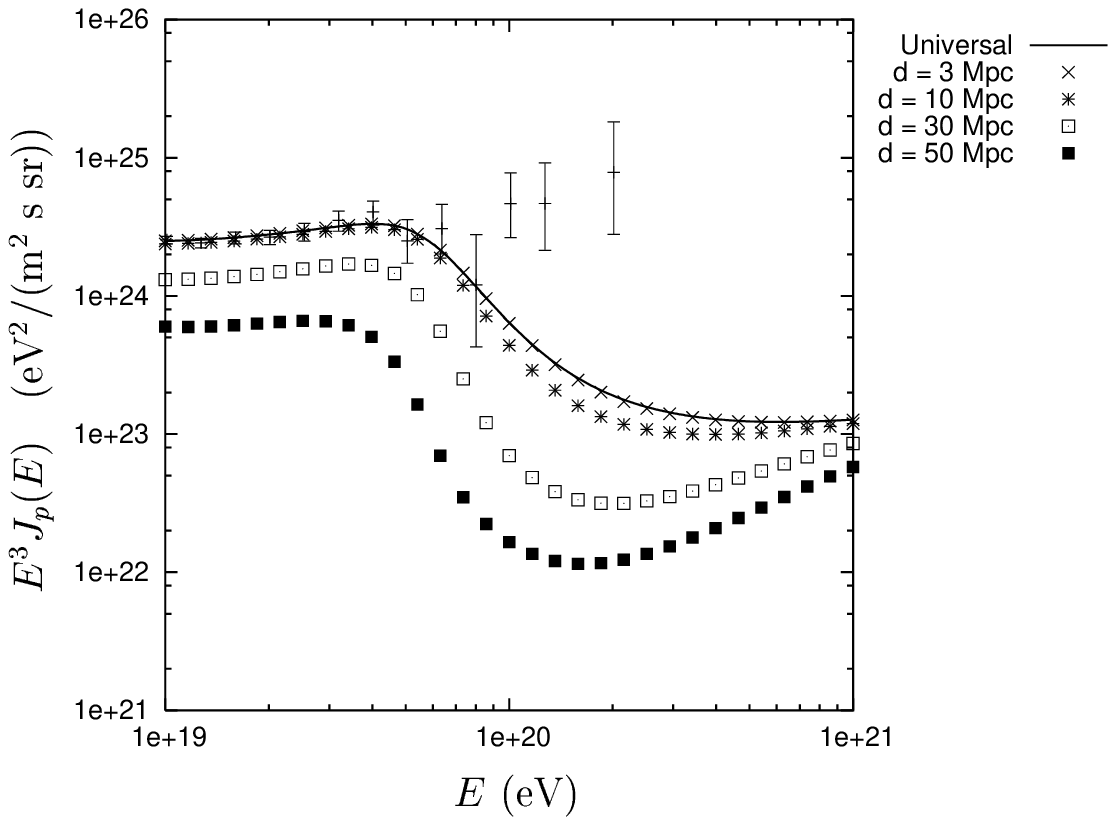}
&
\includegraphics[width=0.45\textwidth]{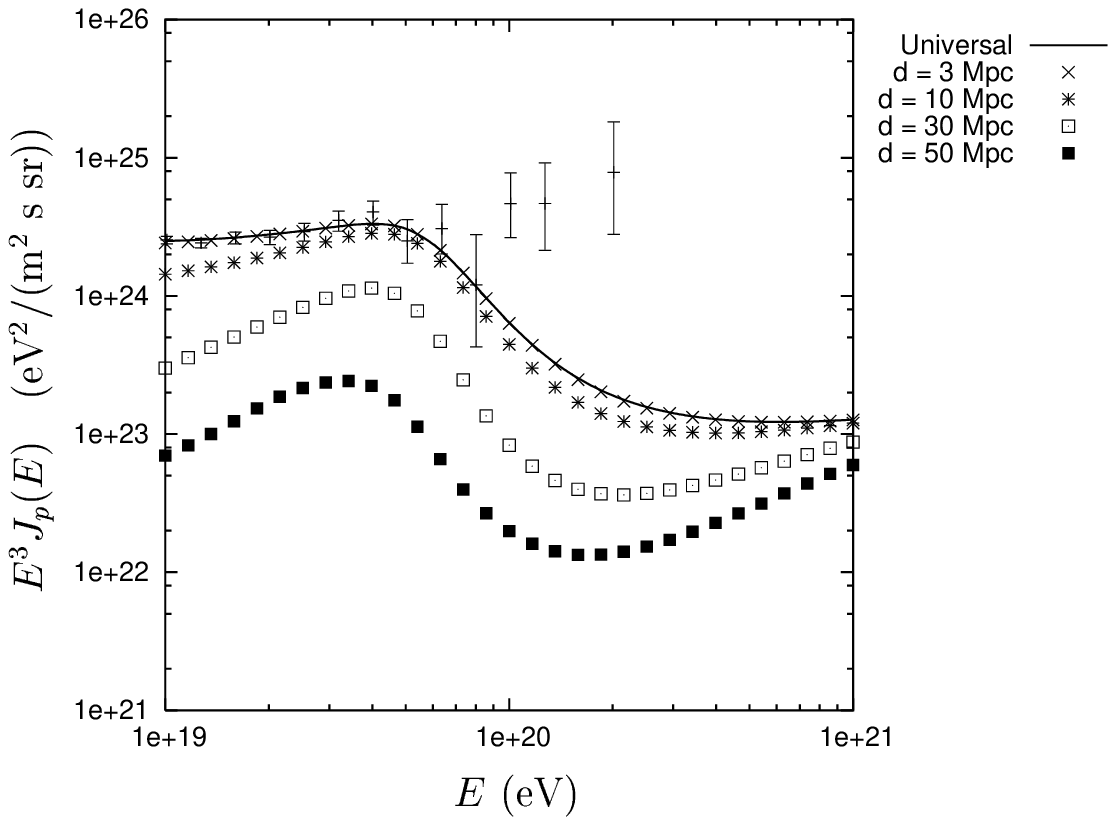}
\\
\end{tabular}
\caption{
Convergence of the diffusion spectrum to the universal spectrum in the
case of the Bohm diffusion,
$D=D_0 (E/E_c)$ at $E<E_c$ (right panel) and in the case of diffusion
with $D=D_0$ at $E<E_c$ (left panel). In the calculations the
complex generation spectrum (\ref{complex}) has been used.
The separations $d$ of the sources are indicated in the figure. The
spectra are compared with the AGASA data.}
\label{fig2}
\end{center}
\end{figure}

In Fig. \ref{fig2} we compare the universal
spectrum with diffusion spectra characterized by different separations $d$.
The diffusion spectra are computed using the diffusion length (\ref{l_d})
for the two extreme cases $\Lambda_d=r_L(E)$ and $\Lambda_d=l_c$, 
both for the representative case $B_0=100$~nG and $l_c=1$~Mpc. 
Here and below we use for the total
lattice size $a=300$~Mpc, while the control
calculations have been performed up to  $a=1000$~Mpc. Increasing $a$
does not change the fluxes.   The universal spectrum
is calculated with time-independent energy losses, as in the Syrovatsky
solution, according to Eq. (\ref{u-spectrum}) with
${\cal L}_0=L_p n_s\simeq 2.8\times 10^{46}$~erg/Mpc$^3$yr.

One can see that in both cases the diffusion spectrum tends to the
universal one as the separation $d$ between sources diminishes.

\section{Diffusive propagation}
\label{diff-par}

In this section we will study the propagation of UHE particles in the
intergalactic turbulent magnetic plasma. Turbulent motion can be considered
as random pulsations at different scales, described as an ensemble of
many waves with large amplitudes and random phases. The energy is
assumed to be injected at the largest scale $l_c$ due to the external
processes, such as the contraction of large-scale structures, large-scale
shocks etc, and the scale $l_c$ might be determined by these processes.
This scale might coincide with the size of the region filled by the 
turbulent plasma, in case of small structures. The smallest
scale $l_{\rm min}$ is determined by the dissipation of turbulent motion
into thermal energy.

We shall consider {\em magnetic turbulence} described by the
superposition of MHD waves $A\exp [ i(-\omega t+ \vec{k}\vec{r})]$
with different frequencies $\omega$ and different wave numbers
$\vec{k}$. The amplitude $A$ is related to magnetic field $B$,
density $\rho$, pressure $P$, velocity $v$ etc. These waves
propagate with the Alfven velocity $u_A=B/\sqrt{4\pi\rho}$.
The kinetic energy of gas in turbulent motion and the magnetic energy in
MHD waves are equal at all scales $\lambda$ (Landau and Lifshitz (1987)):
\begin{equation}
\rho v_{\lambda}^2 \sim B_{\lambda}^2/4\pi.
\label{equip}
\end{equation}
For a detailed description of magnetic turbulent plasma and its
description as an ensemble of MHD waves we refer the reader to the
books: Landau and Lifshitz (1987), Arzimovich and Sagdeev (1979).

For practical calculations of the UHECR propagation, the magnetic
field scale spectrum $\langle B^2 \rangle_k$ as a function of $k$ is needed.
One should clearly distinguish between the magnetic
field $\vec{B}_{\lambda}=\langle \vec{B} \rangle_{\lambda}$,
on the scale $\lambda=2\pi/k$ and the Fourier component of the field
$\vec{B}(k)$ (see Appendix). The physical magnetic field which
determines the diffusion of UHE particles is $B_{\lambda}$.

The turbulence spectrum is described by the dependence of the spectral energy
density $w(k)$ on the wave number $k$. It can be expressed using the
magnetic energy density $kw(k)\sim B_{\lambda}^2/8\pi$. Usually
the spectra have a power-law form: $w(k) \propto k^{-m}$.

In many practical cases the observations confirm the Kolmogorov
(Kolmogorov (1941)) spectrum of turbulence with $m=5/3$. This
spectrum belongs to a class where the initial energy is assumed to be
injected at the largest scale $l_c$ and then is transferred to
lower scales $l$ in non-linear processes of wave interactions
until it dissipates to thermal energy at the lowest scale $l_{\rm
min}$. The Kolmogorov spectrum can be derived assuming that energy
flux from one scale to another does not depend on the scale $k$
(Arzimovich and Sagdeev (1979)), i.e.
\begin{equation}
kw(k)/\tau_k=const,
\label{Eflux}
\end{equation}
where $\tau_k$ is the time for energy transfer to the scale $k$.
Since this process is caused by the non-linear term $(v\nabla)v \sim kv^2$
in the Euler equation, one can estimate $\tau_k$ as
\begin{equation}
\tau_k \sim 1/kv_k,
\label{tau_k}
\end{equation}
where the turbulent velocity can be estimated from equation
$kw(k) \sim \rho v_k^2$. Then one immediately obtains
$w(k) \propto k^{-5/3}$, i.e. the Kolmogorov spectrum.

Landau and Lifshitz (1987) argue that while
scale-independent energy flux is a reasonable assumption in
ordinary hydrodynamics, for MHD waves it is proportional to
$v_k^4$. In this case one obtains the Kraichnan (1965) spectrum
$w(k) \propto k^{-3/2}$.

In the case of shock waves, the spectrum is (Vainstein et al (1989))
$w(k) \propto k^{2}$.

\subsection{Diffusion coefficient for UHE protons}

Now we shall proceed with calculations of the diffusion coefficient
$D(E)$ for UHE protons in extragalactic magnetic fields, which are
characterized by the spectrum of turbulence $w(k) \propto k^{-m}$,
with the basic scale $l_c$ and with the magnetic field on this scale
$B_0$. We shall use the characteristic energy $E_c$, defined by
Eq. (\ref{E_c}) from the condition $r_L(E_c)=l_c$, where the Larmor
radius $r_L$ is given by Eq. (\ref{r_L}).

First we obtain the asymptotic expression for $D(E)$ valid when
$r_L \gg l_c$, i.e. at $E \gg E_c$. The diffusion length $l_d$ is determined
as the distance at which the average angle of scattering satisfies
$$
\langle \theta^2\rangle \sim n\left (\frac{l_c}{r_L} \right)^2 \sim 1,
$$
where the number of scatterings $n$ is given by $n \sim l_d/l_c$.
It results in the diffusion length $l_d(E)\approx r_L^2/l_c$, and thus the
diffusion coefficient, $D(E)=(1/3)cl_d(E)$, in the asymptotic limit
$r_L \gg l_c$ is given by
\be
D_{\rm asymp}(E)=3.6\times 10^{34}E_{20}^2\left (\frac{100~{\rm nG}}
{B_0}\right )^2\left (\frac{1~{\rm Mpc}}{l_c}\right )~{\rm cm}^2/{\rm s}.
\label{D_asymp}
\ee

We shall calculate now the diffusion coefficient in the low-energy
limit $r_L \ll l_c$ (i.e. at $E\ll E_c$). We will follow the V.~Ptuskin
method given in Chapter 9 of the book Berezinsky et al. (1990b). Let us
consider the scattering of UHE protons with $r_L \ll l_c$ in a
magnetic field of MHD wave $B_{\lambda}\exp [ i(\vec{k}\vec{r}-\omega t)]$. 
The magnetic field on the basic scale
$\vec{B}_0$ is considered as constant field for smaller scales
$\lambda$.

Scattering of UHE protons off the MHD waves in the regime $r_L \ll
l_c$ is dominated by resonance scattering (see Lifshitz and
Pitaevskii (2001), Sections 55 and 61). The condition for resonant
scattering is given by $\omega'=s \omega'_B$, where $\omega'$
and $ \omega'_B = eB/\gamma'mc$ are the wave frequency and
the giro-frequency in the system $K'$ at rest  with particle motion
along the field $\vec{B_0}$, and $s=0,~\pm 1,~\pm 2 ...$. After
a Lorentz transformation to the laboratory system, one obtains
\begin{equation}
\omega - k_{\parallel}v_z=s\omega_B,
\label{res}
\end{equation}
where $\omega$ is the wave frequency, $k_{\parallel}$ is the
projection of the wave vector onto the direction of $\vec{B_0}$ ,
$v_z$ is the projection of particle velocity onto the same
direction, and $\omega_B=eB/\gamma mc$. For a magnetized plasma
$r_L \ll \lambda_{\perp}$ (or equivalently
$k_{\perp}v_{\perp}/\omega_B \ll 1$) and $s= \pm 1$ (Berezinsky et al.
(1990b)). Thus from Eq. (\ref{res}) one derives the resonant wave
number
\begin{equation}
k_{\parallel}^{\rm res}= \left | \frac{\omega(k) \pm \omega_B}{v_z}
\right | \approx
\frac {\omega_B}{v \mu}=\frac{1}{r_L\mu},
\label{k_res}
\end{equation}
where $\mu=\cos\theta$ and $r_L=v/\omega_B$ is the Larmor radius.
In deriving Eq. (\ref{k_res}) we have used $\omega(k) \ll
\omega_B$, which follows from the dispersion relation for Alfven
waves $\omega(k)=u_A k_{\parallel}$ (Landau and Lifshitz (1987)),
and $v/u_A \gg 1$ for ultrarelativistic particles.

We shall assume here first the Kolmogorov spectrum normalized at
the basic scale $k_0=2\pi/\lambda_0$:
\begin{equation}
k w(k)=k_0 w_0 (k/k_0)^{-2/3},
\label{kw_k}
\end{equation}
where $w(k)$ is the spectral magnetic energy density normalized
as $k_0 w_0= B_0^2/8\pi$.

We shall adopt from Berezinsky et al. (1990b) (chapter 9) the
diffusion coefficient $D_{\parallel}$ (in the direction of
$\vec{B}_0$) expressed with the frequencies of particle scattering
off the waves:
$$
\nu(\mu,k_{\rm res})=\frac{1}{2}(\nu^++\nu^-)=2\pi^2\omega_B k_{\rm res}
\omega(k_{\rm res})/B_0^2
$$
\begin{equation}
D=\frac{v^2}{4}\int_0^1 d\mu \frac{1-\mu^2}{\nu(\mu,k)},
\label{D}
\end{equation}
where $\nu^+$ and $\nu^-$ correspond to waves propagating along the
field and in the opposite direction, respectively. From now on the
subscript $\parallel$ will be omitted.

Using the formulae above, and performing integration over $\mu$ we
obtain
\begin{equation}
D(E)= \frac{18}{7\pi(2\pi)^{2/3}}vl_c \left ( \frac{r_L}{l_c}\right )^{1/3}
\label{D_LE}.
\end{equation}
Numerically
\begin{equation}
D(E)= 2.3\times 10^{34}E_{20}^{1/3}
\left (\frac{100~{\rm nG}}{B_0}\right )^{1/3}
\left ( \frac{l_c}{1~{\rm Mpc}} \right )^{2/3}~{\rm cm}^2/{\rm s}
\label{D-num}.
\end{equation}
$D(E)$ from Eq. (\ref{D-num}) and from Eq. (\ref{D_asymp}) match
together fairly well: at energy $E_c$ they differ by $40\%$.

The calculations for other spectra are similar: in case of
the Kraichnan spectrum $w(k) \propto k^{-3/2}$, one obtains
$D(E) \propto E^{1/2}$, and for diffusion on shock waves,
$w(k) \propto k^{-2}$,~  $D=const$ follows. \\*[2mm]
{\em Diffusion coefficient for static magnetic field}.\\
This case can be considered as scattering off MHD waves in
the limit $\omega(k) \to 0$. Assuming the spectrum
$$
kw(k)= \frac{B_0^2}{8\pi}\left ( \frac{k}{k_0}\right )^{-\alpha},
$$
with $\alpha <1$ for convergence of the integral, we obtain,
performing the same calculations as above:
\begin{equation}
D(E)= \frac{2(2\pi)^{-\alpha}}{\pi (1-\alpha)(3-\alpha)} l_c v
\left ( \frac{r_L}{l_c} \right )^{1-\alpha}~.
\label{static}
\end{equation}

\begin{figure}[t!]
\begin{center}
\begin{tabular}{ll}
\includegraphics[width=0.45\textwidth]{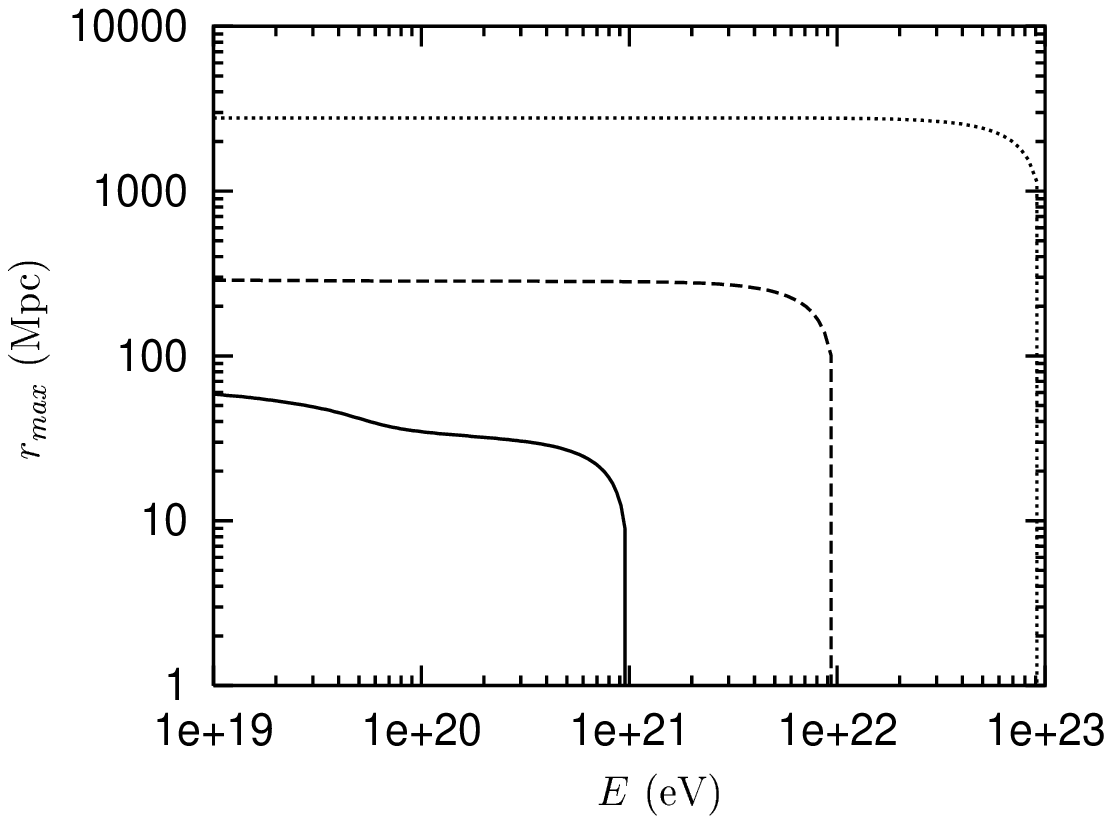}
&
\includegraphics[width=0.45\textwidth]{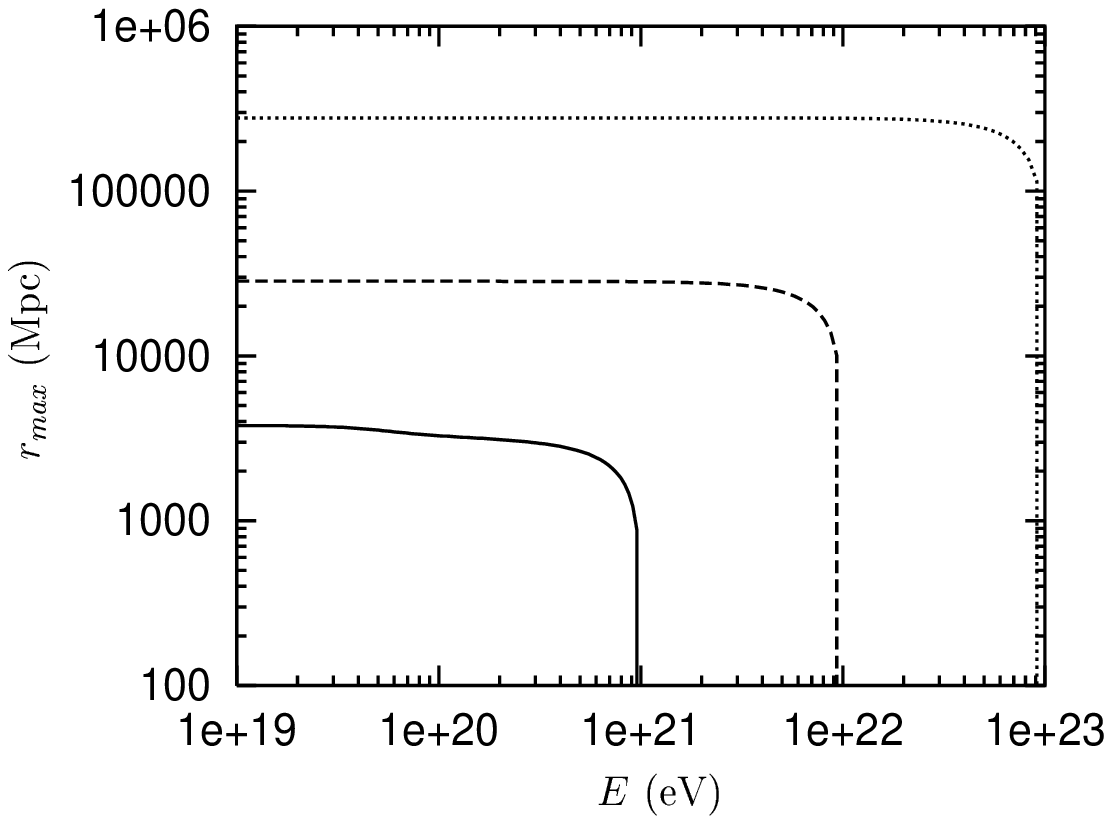}
\\
\end{tabular}
\caption{
Maximum  distance to a source
$r_{\rm max}(E)=2\sqrt{\lambda(E,E_g^{\rm max})}$
for diffusive propagation. Continous line:~
$E_g^{\rm max}=1\times 10^{21}$ eV,~ dashed line:~
$E_g^{\rm max}=1\times 10^{22}$ eV,~
dotted line:~ $E_g^{\rm max}=10^{23}$ eV. Magnetic field configuration is
$(B_0,l_c)=(100~{\rm nG},~ 1~ {\rm Mpc})$ (left panel) and
$(B_0,l_c)=(1~{\rm nG},~ 1~ {\rm Mpc})$ (right panel).
}
\label{fig3}
\end{center}
\end{figure}

\subsection{Intergalactic medium and formation of turbulent spectrum}

The intergalactic medium is represented by different structures
(clusters of galaxies, superclusters, filaments and voids), where
the medium properties are very much different. In the estimates
below we shall normalize all quantities by the baryon density
$n_b=2.75\times 10^{-7}$ cm$^{-3}$, corresponding to
$\Omega_b=0.044$ from WMAP measurements (Spergel et al. (2003)),
and by the temperature $T \sim 10^6$ K (see e.g. the simulation of
Dav\'e et al. (2001) for filaments). Hence, we adopt the sound
speed
\begin{equation}
c_s=(\gamma T/m_H)^{1/2}=1.2 \times 10^7 (T/10^6)^{1/2} {\rm cm/s},
\label{c_s}
\end{equation}
the Alfven velocity
\begin{equation}
u_A= \frac{B}{\sqrt{4\pi\rho_b}}= 4.2\times 10^5 \frac{B}{1~{\rm
nG}} \left (\frac{2.75\times 10^{-7}~{\rm
cm}^{-3}}{n_b}\right)^{1/2}~ {\rm cm/s}, \label{u_A}
\end{equation}
and the  Coulomb scattering length for electron-electron and proton-proton
scattering
\begin{equation}
l_{\rm sc}=\frac{T^2}{4\pi e^4 n_b L}= 1.7 \left (\frac{T}{10^6}\right )^2
\frac{2.75\times 10^{-7}}{n_b}~{\rm kpc},
\label{l_sc}
\end{equation}
where $L \sim 20$ is the Coulomb logarithm. The short scattering
length $l_{\rm sc}$ in comparison with the basic scale $l_c$ is one of
the conditions which are needed to provide the turbulent regime.

With the characteristics of the media above, we shall discuss now,
whether the equilibrium turbulence spectrum in the gas can be formed during
the age of the universe $t_0$. We shall make estimates for the
Kolmogorov spectrum.

The relaxation time $\tau_k$ to the equilibrium Kolmogorov spectrum is
given by Eq. (\ref{tau_k}) as $\tau_k \sim 1/k v_k$, where $v_k$ is
the turbulent velocity on the scale $k$. Taking  $v_k$ from
$\rho v_k^2 \sim 2kw(k)$, and using the Kolmogorov spectrum normalized at
the basic scale as in Eq. (\ref{kw_k}), we obtain
\begin{equation}
\tau_k= \frac{1}{k_0}\left (\frac{\rho}{2k_0w_0}\right )^{1/2}
\left ( \frac{k}{k_0} \right )^{-2/3}.
\end{equation}
The longest time is needed for the formation of the spectrum at the largest
scale $k_0$:
\begin{equation}
\tau_0=\frac{1}{k_0}\left (\frac{\rho}{2k_0w_0}\right )^{1/2}.
\end{equation}
Using $k_0w_0=B_0^2/8\pi$ and $v_0= B_0/\sqrt{4\pi\rho}=u_A$, we
obtain
\begin{equation}
\tau_0 \sim l_c/u_A.
\end{equation}
Numerically it gives
\begin{equation}
\tau_0=2.4\times 10^{11}\frac{l_c}{1~{\rm Mpc}}\frac{1~{\rm nG}}{B_0}
\left (\frac{n_b}{2.75\times 10^{-7}}\right )^{1/2}~{\rm yr}.
\label{tau-num}
\end{equation}

Another estimate of the relaxation time $\tau_0$ can be obtained for
the sonic turbulence with the Kolmogorov spectrum. The shortest time
is given by $\tau_0=l_c/c_s$, which for a turbulent plasma is the analogue
of the causality condition. Numerically it gives
\begin{equation}
\tau_0=0.84\times 10^{10}\frac{l_c}{1~{\rm Mpc}}
\left (\frac{10^6~K}{T}\right )^{1/2}~{\rm yr}.
\label{sonic}
\end{equation}
For a more accurate estimate one can use the relaxation time for the
sonic turbulence from Arzimovich and Sagdeev (1979)
\begin{equation}
\tau_k \sim \frac{n_b T}{kw_k}\frac{1}{\omega_k}.
\end{equation}
Using the dispersion relation $\omega_k=k(T/m)^{1/2}$ and the
Kolmogorov spectrum $kw_k =k_0w_0(k/k_0)^{-2/3}$, with $k_0 w_0 \sim nT$,
we obtain numerically an estimate close to that given by Eq. (\ref{sonic}).

From the above estimates we can conclude that the Kolmogorov spectrum cannot
be reached for MHD turbulence in the voids, where the magnetic field is 
presumably  small, $B_0 \leq 1$~nG. However,
it can be established in filaments (see Eq.(\ref{tau-num})) in case 
$B_0 > 1$~nG and $l_c \leq 1$~Mpc, and it has enough
time to be developed in galaxy clusters, where the magnetic field is strong and
density of gas is larger than in other structures. For other types of
turbulence, e.g. for sonic turbulence, these conditions can be somewhat
relaxed.

\subsection{Some features of diffusive propagation}

Inspired by numerical simulations (Yoshiguchi et al. (2003)) we
study the diffusion in various magnetic field configurations with the
basic parameters $(B_0,l_c)$ in the intervals $B_0=10 - 1000$~nG
and $l_c=1 - 10$~Mpc. As a representative configuration we
consider (100~nG,~1~Mpc) with $E_c \approx 1\times 10^{20}$~eV.

The calculated diffuse energy spectra are characterized by the
sets $(B_0,l_c,d)$, where $d$ is the separation between sources.
We shall use the following definitions. The case when at all
relevant (observed) energies $d > l_d(E)$ (or $d\gg l_d(E)$)  we
shall refer to as {\em diffusion in a strong magnetic field}. The
case when $d \lsim l_d(E)$ corresponds to {\em diffusion in a weak
magnetic field}.\footnote{Note, that this definition does not
guarantee the (quasi)rectilinear propagation of particles.}
The extreme case $d \ll l_d(E)$ results in the universal spectrum.

\noindent
{\em Minimal distance to a source.}\\
For the diffusion approximation to be valid, a source must be at a distance
$r>r_{\rm min}(E)$. In principle, velocity of light does not enter the
diffusion equation, but this equation is not valid when the propagation
velocity exceeds the light speed $c$. The value of $r_{\rm min}(E)$ can be
estimated from the condition that the diffusive propagation time must be
longer than the time of rectilinear propagation,
$$
t_{\rm prop}\sim \frac{r^2}{D(E)} > \frac{r}{c},
$$
as
\be
r_{\rm min}(E)\sim \frac{1}{3} l_d(E) .
\label{r_min}
\ee
For a source at distance $r\leq r_{\rm min}(E)$ the protons with 
energy $E$ or higher propagate in a quasi-rectilinear regime.
At $E \geq E_c$ one has
$$ r_{\rm min}(E)=\frac{1}{3} \left (\frac{E}{E_c} \right)^2 l_c. $$

\noindent
{\em Maximal distance to a source.}\\
As seen from Eq. (\ref{syr-sol}) the contribution of a source to the flux at
energy $E$ becomes negligible at distances $r>r_{\rm max}(E)$ with
\be
r_{\rm max}(E)=2\sqrt{\lambda(E,E_g^{\rm max})},
\label{r_max}
\ee
where $E_g^{\rm max}$ is the maximum generation energy provided by a source.

For the representative case $(B_0,l_c)=(100~ {\rm nG},1~ {\rm Mpc})$,
$r_{max}$ is plotted in Fig. \ref{fig3} as a function of
$E$ for different values of $E_g^{max}$. One can see that in the case
of diffusive propagation only nearby sources contribute the UHECR flux.

\begin{figure}[t!]
\begin{center}
\includegraphics[width=0.5\textwidth]{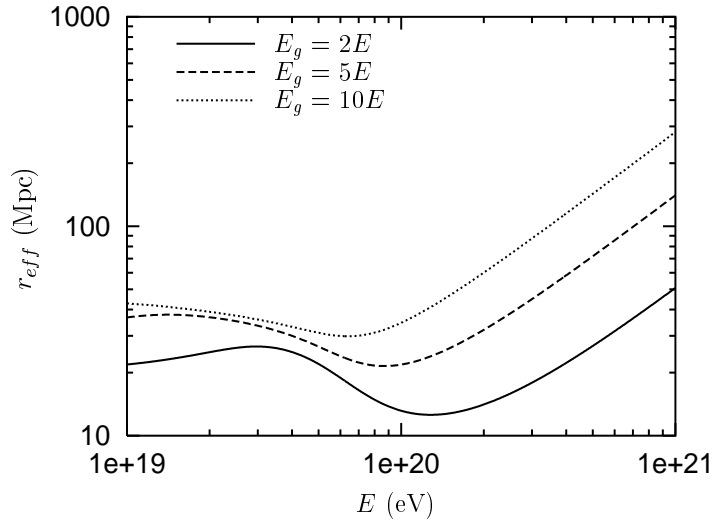}
\caption{Effective distance to a source
$r_{eff}(E)=2\sqrt{\lambda(E,E_g/E)}$ with magnetic field
configuration $(B_0,l_c)=(100~ {\rm nG},~ 1~{\rm Mpc})$. }
\label{fig4}
\end{center}
\end{figure}

In Fig. \ref{fig4} we have plotted also
$$r_{\rm eff}(E)=2\sqrt{\lambda(E,E_g/E)} $$
for different (fixed) ratios $E_g/E=2, 5$ and 10. The increase of
$r_{\rm eff}$ with $E$ provides the increase of diffuse flux $J_p(E)$ with
energy when the distance between sources $d$ is large enough 
(see Fig.~\ref{fig2}).

\section{Diffuse fluxes in the diffusion approximation}
\label{diff_diff}

In this Section
we will compute the diffuse flux according to Eq.~(\ref{Sy-diffuse})
as the sum of fluxes from single sources located in the lattice vertexes
with a separation $d$ and with a total size $a$. We will
assume the complex generation spectrum given by Eq.~(\ref{complex}).
For the diffusion coefficient we use $D(E)=(1/3)cl_d(E)$
with $l_d(E)$ given by the interpolation formula (\ref{l_d}) for the case of
Bohm diffusion at $E<E_c$.

At small distances $r \leq r_{\rm min}(E)$, given by Eq. (\ref{r_min}),
the fluxes from the individual sources are calculated in the 
rectilinear approximation, and the diffuse flux is given by
\be
\label{diff-rect}
J_p^{\rm rect}(E)=\frac{L_p K(\gamma_g)}{4\pi} \sum_i
\frac{q_{gen}(E_g(E,r_i))}{r_i^2} \frac{d E_g(E,r)}{d E}
\ee

\begin{figure}[t!]
\begin{center}
\includegraphics[width=0.5\textwidth]{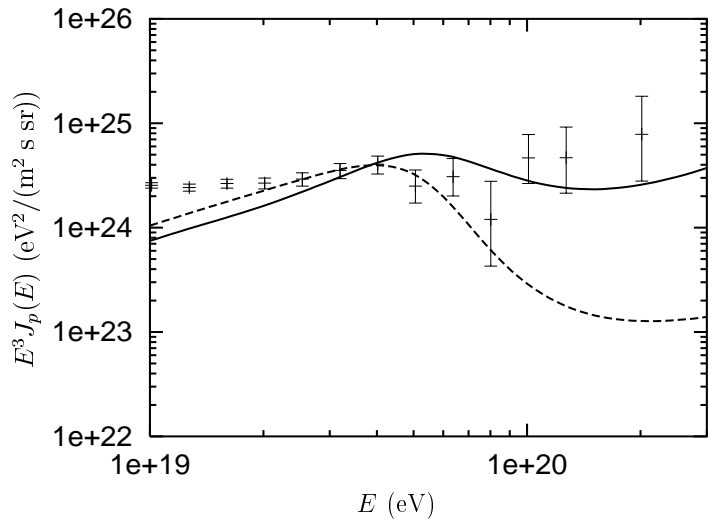}
\caption{
Diffusion spectra for two different configurations $(B_0,~l_c,~d)$. 
Dashed line corresponds to
(100~nG,~1~Mpc,~30~Mpc) with source luminosity 
$L_p=8.5\times 10^{43}$~erg/s, while continuous line is shown for the same
$d$ but for much stronger magnetic field: (1000~nG,~1~Mpc,~30~Mpc),
with source luminosity $L_p=1.5\times 10^{47}$~erg/s. In the latter
case the GZK cutoff is weak.
The spectra are compared with the AGASA data.
}
\label{fig5}
\end{center}
\end{figure}

At large distances $r \geq r_{\rm min}(E)$ the diffuse flux is given by
Eq. (\ref{Sy-diffuse}), and with explicit normalization has the form
\be
J_p^{diff}(E)=\frac{c}{4\pi}
\frac{L_p K(\gamma_g)}{b(E)} \sum_i \int_{E}^{E_{max}}
dE_g q_{gen}(E_g) \frac{exp\left [-\frac{r_i^2}{4\lambda(E,E_g)} \right ]}
{\left ( 4\pi\lambda(E,E_g)\right )^{3/2}},
\ee
where $E_{max}$ is the maximum acceleration energy
(see Section~\ref{syrovatsky}).
The calculated spectra are presented in Fig. \ref{fig5}.

In Fig. \ref{fig5} the diffuse spectra are shown for the complex
generation spectrum (\ref{complex}) and for two sets of $(B_0,l_c,d)$
equal to (100~nG,~1~Mpc,~30~Mpc), shown by dashed curve,  and
(1000~nG,~1~Mpc,~30~Mpc), shown by solid curve. Both of them are
characterized by the same $d=30$~Mpc, but in the latter case the
magnetic field is much  stronger. This case is characterized by  
$d\gg l_d(E)$
at all observable energies, and hence corresponds to diffusion in
the strong magnetic field (see Section \ref{diff-par}). From
Fig. \ref{fig5} one can observe that GZK cutoff is weak in this case.

The dashed curve in Fig. \ref{fig5} is characterized by 
$l_d(E)= 1 (E/E_c)^2$~Mpc at $E>E_c=1\times 10^{20}$~eV,
and this regime of propagation can be described
as an intermediate one between that for strong and weak magnetic fields.

The regime with weak GZK cutoff
(propagation in a strong magnetic field) requires a much higher luminosity
$L_p$ to fit the observational data, $L_p=1.5\times 10^{47}$~erg/s, while
the intermediate case (100~nG,~1~Mpc,~30~Mpc) needs only
$L_p=8.5 \times 10^{43}$~erg/s.

We shall analyze now the regime of propagation in strong magnetic
field in more detail. First we will comment qualitatively why   
in the diffusion approximation the GZK cutoff
in the spectrum might be weak or absent.

The essence of the GZK cutoff consists of much different energy losses 
above and below $3\times 10^{19}$~eV. Consider, for example, two protons
with energies $1\times 10^{19}$~eV and $1\times 10^{20}$~eV, both propagating
rectilinearly from a remote single source. The first proton looses little
energy: the ratio $E_g/E$ (with $E_g$ being the generation energy) is 
not large, and the flux of these protons is weakly suppressed by the 
generation spectrum.
The energy losses of the second proton are large, $E_g/E$ is very high and
the flux suppression is dramatic (the GZK cutoff). 
Now consider the diffusive propagation. 
A proton with energy $1\times 10^{19}$~eV, because of the $D(E)$ dependence, 
travels much longer time than a proton with $1\times 10^{20}$~eV, 
and it results in increased ratio $E_g/E$, making this ratio 
comparable with that for the second proton. It causes a less steep 
GZK cutoff, or its absence. However, the price for the absence of 
the GZK cutoff is a very high luminosity of the sources $L_p$, needed to 
provide the observed flux at $E \gsim 1\times 10^{19}$~eV, e.g.
$L_p \gsim 1\times 10^{47}$~erg/s for the spectrum shown in Fig.~\ref{fig5}.

\begin{figure}[t!]
\begin{center}
\includegraphics[width=0.5\textwidth]{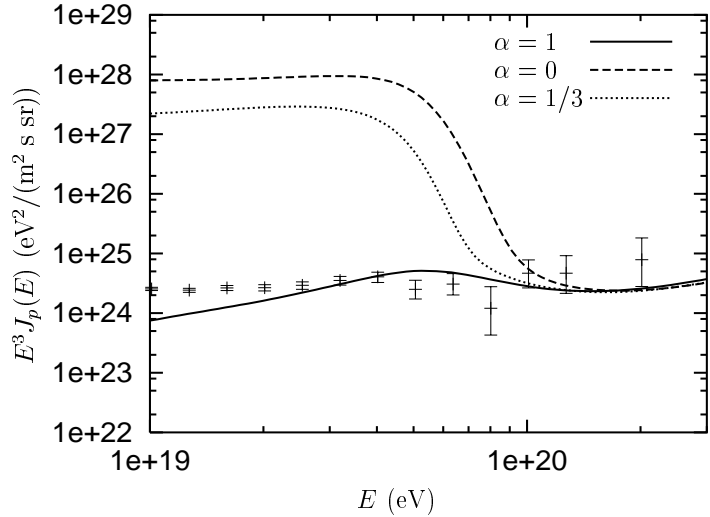}
\caption{ Diffuse fluxes from lattice-distributed sources with 
$d=30$~Mpc and for magnetic configuration (1000~nG,~1~Mpc). The cases 
$\alpha= 0,~1/3,~1$ correspond to $D=const$, Kolmogorov and Bohm 
diffusion, respectively at $E < E_c \approx 1\times 10^{21}$~eV.
Luminosity of a source is $L_p=1.5\times 10^{47}$ erg/s.  
}
\label{fig6}
\end{center}
\end{figure}

Let us now come over to the quantitative analysis of absence of 
the GZK cutoff in the strong magnetic fields. Consider the three cases 
of diffusion regimes at $E < E_c$ with $\alpha= 0,~1/3,~1$, i.e. $D=const$,
Kolmogorov and Bohm regimes, respectively (see Eq.~(\ref{alpha})). 
The diffuse spectra calculated for configuration (1000~nG,~1~Mpc,~30~Mpc)
are displayed in Fig.~\ref{fig6}

Why the three spectra are so much different at low energies and are 
the same at high energies?

The quantitative explanation can be given explicitly in terms of
$\lambda(E,E_g)$, which is the basic parameter of the Syrovatsky solution 
(see Eq.~(\ref{syr-sol})). 

\begin{figure}[t!]
\begin{center}
\begin{tabular}{ll}
\includegraphics[width=0.45\textwidth]{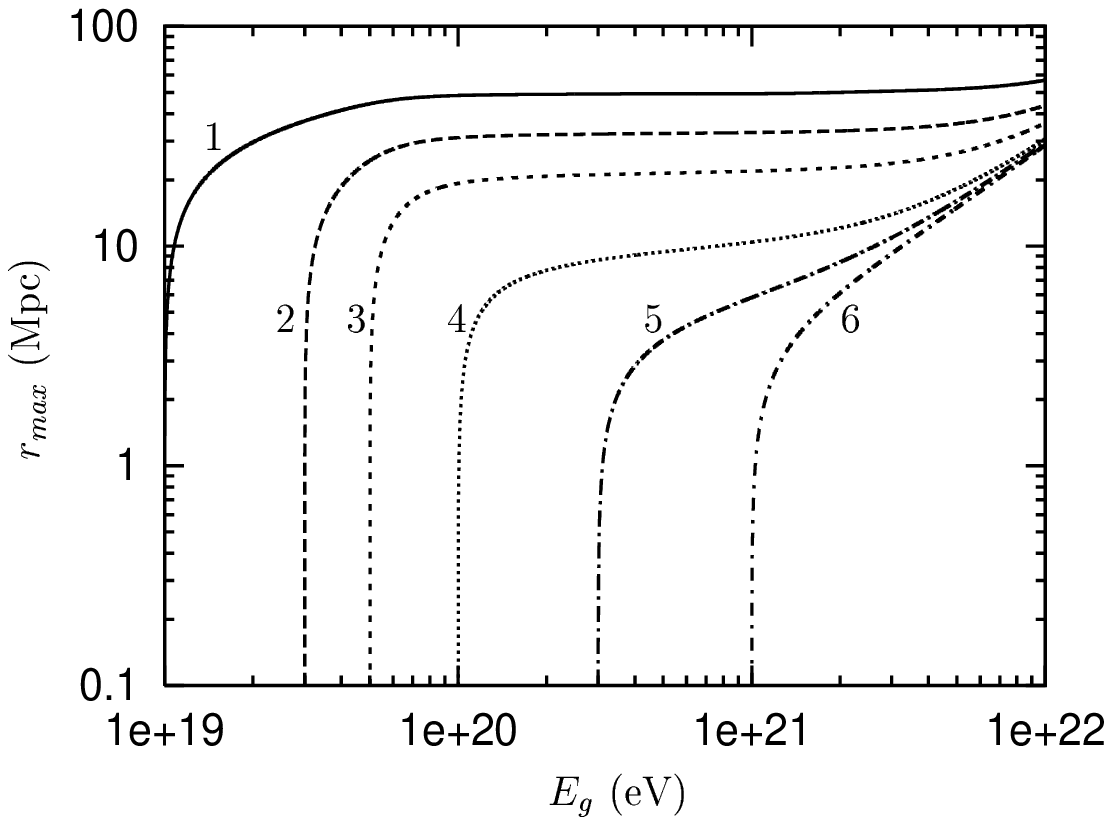}
&
\includegraphics[width=0.45\textwidth]{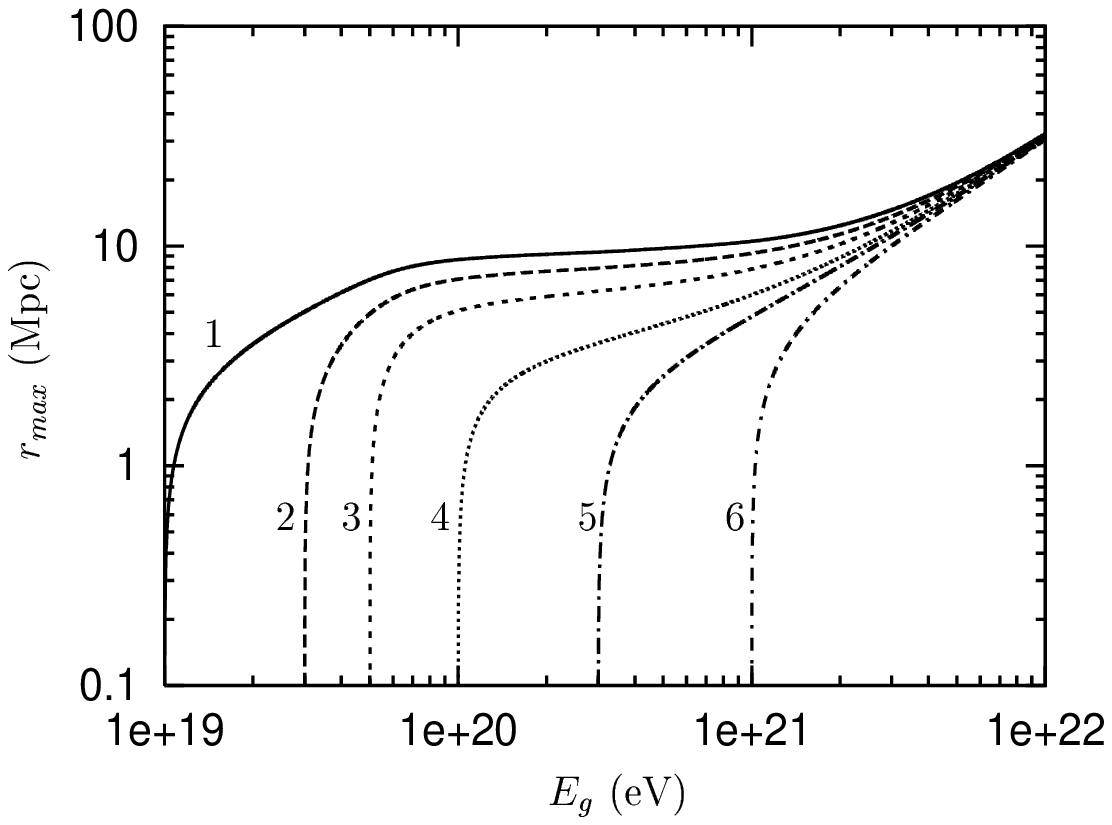}
\\
\end{tabular}
\caption{The values of $r_{\rm max}(E,E_g)=2\sqrt{\lambda(E,E_g)}$ as 
function of 
generation energy $E_g$ for the $D=const$ diffusion (left panel) and
the Bohm diffusion (right panel). The magnetic field configuration 
is (1000~nG,~ 1~Mpc,~ 30~Mpc). The observed energies E are 
$1\times 10^{19}$~eV - curve 1,~ $3\times 10^{19}$~eV - curve 2,~ 
$5\times 10^{19}$~eV - curve 3,~ $1\times 10^{20}$~eV - curve 4,~ 
$3\times 10^{20}$~eV -curve 5,~ $1\times 10^{21}$~eV - curve 6.
} 
\label{fig7}
\end{center}
\end{figure}

In Fig.~\ref{fig7} we plot the values of 
$$
r_{\rm max}(E,E_g)=2\sqrt{\lambda(E,E_g)},
$$
which according to Eq.~(\ref{syr-sol}) determines the maximum distance
to a source in case the observed energy is $E$ and generation energy 
is $E_g$. The left panel of Fig.~\ref{fig7} corresponds to 
$D=const$, the right panel -- to the Bohm diffusion.   

From Fig.~\ref{fig7} one can see that in the energy interval 
$(1 - 4)\times 10^{19}$~eV~~ $\lambda (E,E_g)$ practically does not
depend on $E_g$. Then from Eq.~(\ref{syr-sol}) one obtains for a
single source   
\begin{equation}
n_p(r,E)=\frac{1}{(\gamma_g-1)(4\pi)^{3/2}}\frac{Q(E)}
{(\lambda(E))^{3/2}\beta(E)}\exp \left( -\frac{r^2}{4\lambda(E)} \right ),
\label{lowE}
\end{equation}
where $\beta(E)=E^{-1}dE/dt$.

From Eq.~(\ref{lowE}) and values of $\lambda(E)$ for the two 
diffusion regimes, one may observe the main effect : strong
suppression of flux from nearby sources ($r\sim d \sim 30$~Mpc) in
case of the Bohm diffusion and weak suppression in case of $D=const$ 
diffusion. It occurs because the Bohm diffusion coefficient,
$D_B(E)=D_0(E/E_c)$, is small at $E \ll E_c$, and hence $\lambda(E)$ is
also small, which results in exponential suppression of the flux,
according to Eq.~(\ref{lowE}). In case $D(E)=D_0$, diffusion
coefficient is large, $\lambda(E)$ is large too (see left panel of 
Fig.~\ref{fig7}), and exponential suppression in Eq.~(\ref{lowE}) is
much smaller.
This quantitative explanation agrees with 
the qualitative one, given above: large propagation time, i.e. 
small diffusion coefficient or small $\lambda$, suppresses the 
the flux at $E<E_{\rm GZK}$ to the level of the flux at the highest
energies. 

Let us come over to the higher energies.

One can see from Fig.~\ref{fig7} that at $E \geq 1\times 10^{20}$~eV 
for $D=const$ diffusion and $E \geq 1\times 10^{19}$~eV for the Bohm 
diffusion $\lambda (E)$ is small but increases fast with energy up 
to $r_{\rm max}(E,E_g) \sim 30$~Mpc at $E=E_{\rm max}$. The exponent 
in Eq.~(\ref{syr-sol}) grows very fast and all other quantities can be taken
out of integral at energy $E_g=E_{\rm max}$. After integration one
arrives at analytic expression
\begin{equation}
n_p(r,E)=\frac{Q(E_{\rm max})}{4\pi r^2}\frac{\sqrt{\lambda_0}}{D_0}
\exp{-\frac{r^2}{4\lambda_0}} \frac{2}{\sqrt{\pi}}
\left (\frac{E_{\rm max}}{E}\frac{\beta(E_{\rm max})}{\beta(E)}\right )
\left ( \frac{E_c}{E_{\rm max}}\right )^2,
\label{highE}
\end{equation}
where $\beta(E)=E^{-1}dE/dt$ and $\lambda_0=\lambda (E,E_{\rm max})$,
which in fact does not depend on $E$. From Eq.~(\ref{highE}) one can see
that $n_p(r,E)$ has universal dependence for all diffusion regimes,
provided by $D(E_{\rm max})=D_0(E_{\rm max}/E_c)^2$ for all of them. 
In asymptotically high energy regime (not seen in Fig.~\ref{fig6})
$E^3n_p(E) \propto E^2$.

We found and analysed above the diffusion regime with weak GZK cutoff
for very strong magnetic field $B_0=1000$~nG. Can this effect exist 
for much weaker magnetic field?  We have not found such regimes in any 
realistic cases we studied. 

The strong restriction to existence of these regimes is imposed 
by upper limit to the distance between sources $d$.
This distance cannot be taken arbitrary large.
They are limited by maximum acceleration energy $E_{\rm max}$, which 
we keep here reasonably high $E_{\rm max}=1\times 10^{22}$~eV. For rectilinear
propagation from a source at distance $r$ the proton with observed energy 
$E=1\times 10^{20}$~eV must have at $r=100$~Mpc 
the generation energy $E_g=1\times 10^{22}$~eV, and thus the sharp
cutoff at $E=1\times 10^{20}$~eV is predicted. For $r=50$~Mpc the 
cutoff energy 
energy is $E=3\times 10^{20}$~eV. Assuming $r \sim d$, one obtains 
$d \lesssim 50 - 100$~Mpc for $E_{\rm max}=1\times 10^{22}$~eV.

We have performed many calculations with magnetic fields in the range 
100 - 300 nG, with $l_c$ in a range 1 - 10~Mpc, for all three
difusion regimes $\alpha=1,~1/3,~0$ and with $d$ in a range 
30 - 50~ Mpc. In all cases the spectra expose GZK cutoff.  

On this basis we conclude that in case 
$E_{\rm max} \leq 1\times 10^{22}$~eV, i.e. for $d \lesssim 50$~Mpc,
the {\em diffusion regime} with weak GZK cutoff appears only at 
very strong magnetic fields $B_0 \sim 1000$~nG, and it needs very high
source luminosities $L_p \sim 10^{47}$~erg/s.

\section{Conclusions}
\label{conclusions}

We have performed a formal study of the propagation of UHE particles,
using an analytic approach. We demonstrated that the distance between
sources is a crucial parameter which strongly affects the diffuse
energy spectrum.

We have proved that, for a uniform distribution of sources, when
the separation between them is much smaller than all characteristic
propagation lengths, most notably the diffusion length $l_d(E)$ and
the energy attenuation length $l_{\rm att}(E)$, the diffuse spectrum of
UHECR has a {\em universal} form, independent of the mode of propagation.
This statement has the status of a theorem and is valid for propagation in
strong magnetic fields. The proof is given using particle number
conservation during propagation and also using the kinetic
equation for the propagation of UHE particles. In particular the exact
solution to the kinetic equation (\ref{kin-eq1}) for a {\em homogeneous}
({\em i.e.} continuous and distance-independent) distribution of
the sources gives exactly the same spectrum as in the method using
particle-number conservation. Note that in Eq.(\ref{kin-eq1}) the
diffusion term is absent due to homogeneous distribution of the
sources.

Another proof of the theorem is given using the diffusion equation
(\ref{diff-eq}), and its exact solution (Syrovatskii (1959)) for a
single source in the case of time-independent energy losses. We
calculated the diffuse flux putting the sources at the vertexes of
a big lattice with size $a$ and with separation $d$ between
vertexes (sources). The results of the calculations are shown in
Fig. \ref{fig2}. One can see that when the distance between
sources diminishes from 50~Mpc to 10~Mpc, the spectra converge to
the universal spectrum (full curves) and at $d=3$~Mpc they become
identical to the universal spectrum. This result is confirmed by
analytic calculations. When the separation between sources is
small, summation over the sources can be replaced by an integration
and the corresponding spectrum (\ref{diffuse}) coincides exactly with
the universal spectrum.

In this paper we have studied the diffusive propagation of UHE
particles in intergalactic space, which is considered as a turbulent
magnetic plasma with baryonic gas density $n_b$ and temperature
$T$, and with two basic turbulence scales $l_c$, where external
energy is injected, and $l_{\rm min}$, where turbulent energy is
dissipated. Turbulent motion is considered to be random pulsations at
different scales, described by an ensemble of magneto-hydrodynamic
(MHD) waves $B_{\lambda}\exp [ i(-\omega t + \vec{k}\vec{r})]$ with
different frequencies $\omega$ and different wave numbers
$\vec{k}$. Diffusion arises due to resonant scattering of
particles in magnetic fields of MHD waves. At ``low'' energies,
when the Larmor radius $r_L$ is much smaller than the basic scale
$l_c$, the diffusion coefficient can be calculated, provided the
spectral energy density of the turbulent plasma, $w(k)$, is known
as a function of $k$. Such calculations for the Kolmogorov spectrum
of turbulence result in the diffusion coefficient given by
Eq.(\ref{D-num}).

It is interesting to note, that the diffusion in static magnetic fields
can be calculated by this method as the limiting case, when wave
frequency $\omega \to 0$. The diffusion coefficient for static
magnetic fields is given by Eq.(\ref{static}).

The diffusion coefficient in the high energy limit, when $r_L \gg l_c$,
can be reliably calculated as the process of multiple scattering.
The diffusion coefficient for this extreme case is given by Eq.(\ref{D_asymp}).
The diffusion coefficients in the low energy regime ($r_L\ll l_c$), given by
Eq.(\ref{D-num}), and in the high energy regime ($r_L\gg l_c$), given by
Eq.(\ref{D_asymp}), match each other well.

In practical calculations of the diffuse fluxes we characterize
the magnetic configuration by three parameters ($B_0,~l_c,~d$),
where $B_0$ is the mean magnetic field on the basic scale $l_c$
and $d$ is the separation of sources. We put the sources at the
vertexes of a lattice of total size $a$, for which we typically
used $a= 300$~Mpc. Increasing $a$ does not change the fluxes. 
The fluxes from the individual sources were found as the
Syrovatsky solution (\ref{syr-sol}), and the diffuse flux was found
by summation over all sources, given by Eq.(\ref{Sy-diffuse}). As a
representative case for a strong magnetic field, we considered the
configuration ($B_0,~l_c$)=(100~nG,~1~Mpc).

An important feature of the diffusion model is that the observed diffuse
flux is produced by nearby sources (see Fig. \ref{fig3} and \ref{fig4}).
The maximum distance $r_{\rm max}$ depends strongly on the
maximum acceleration energy $E_{\rm max}$.
For example, for $E_{\rm max}=1\times 10^{21}$~eV
and $(B_0,l_c)=$ (100~nG,~1~Mpc), the maximum distance is less than
70~Mpc, while for $E_{\rm max}=1\times 10^{22}$~eV this distance is
200~Mpc. For smaller magnetic fields these distances are larger
(see Fig. \ref{fig3} right panel). The small radius $r_{\rm max}$ of
the region, which provides the dominant contribution to the observed
diffuse flux of UHECR, imposes a constraint on the diffusion models, which will
become more severe when/if particles with higher energies will be observed.
As an example, let us consider the representative configuration
(100~nG,~1~Mpc) and highest observed energy $E=3\times 10^{20}$~eV.
In this case we have $E_c=1\times 10^{20}$~eV, $l_d(E)\approx 10$~Mpc
and $l_{\rm att}=21$~Mpc. To avoid rectilinear propagation from nearby
sources we should impose the separation $d > l_d(E)\approx 10$~Mpc. For
a maximum generation energy  of $E_{\rm max}=1\times 10^{21}$~eV, the
maximum distance is $r_{\rm max} < 70$~Mpc  and is only marginally 
consistent with $d> 10$~Mpc. The UHECR sources at $r<70$~Mpc with $d>10$~Mpc 
and with maximum acceleration energy $E_{\rm max}\sim 1\times 10^{21}$~eV
tentatively imply AGN, whose luminosities satisfy the energy
requirement for the observed UHECR fluxes in the case of weak magnetic
fields. However, for the case of a weak GZK cutoff (diffusion in strong
magnetic fields and large separation $d$) the required luminosities are
very high $L_p \sim 10^{47}$ erg/s.

\begin{figure}[t!]
\begin{center}
\begin{tabular}{ll}
\includegraphics[width=0.45\textwidth]{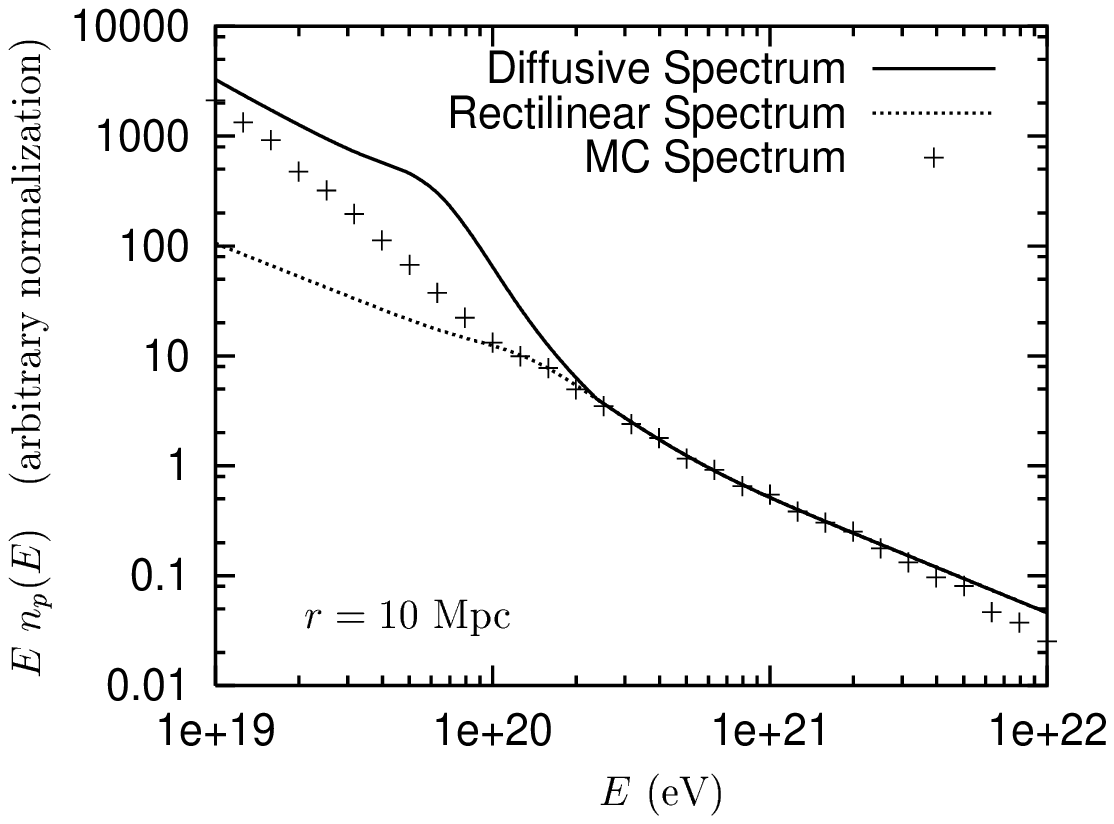}
&
\includegraphics[width=0.45\textwidth]{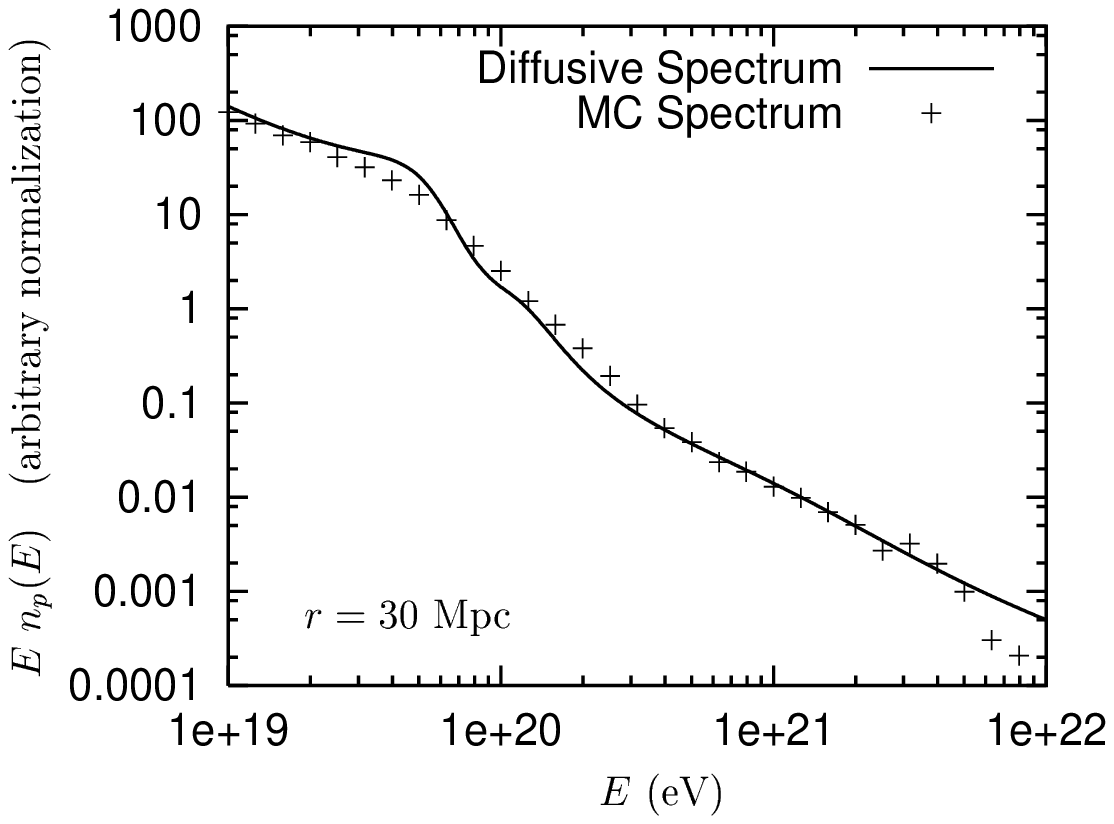}
\\
\end{tabular}
\caption{ Comparison of the analytic  diffusive spectrum with the Monte
Carlo simulation by Yoshiguchi et al. (2003). The
spectra are given in the case of Kolmogorov diffusion with
$(B_0,l_c)=(100~ {\rm nG},1~{\rm Mpc})$ and for a single source
placed at distance $10$~Mpc (left panel) and $30$~Mpc (right
panel). } 
\label{fig8}
\end{center}
\end{figure}

The calculated diffuse energy spectra are shown in Fig. \ref{fig5}
for the Bohm  diffusion coefficient and for the complex generation 
energy spectrum (\ref{complex}).
In case of configuration with a very strong magnetic gield  
(1000~nG,~1~Mpc,~30~Mpc) the spectrum has a weak GZK cutoff 
(the solid curve in  Fig. \ref{fig5}). This spectrum agrees reasonably
with the AGASA excess, but requires very large source luminosity  
$L_p=1.5\times 10^{47}$ erg/s. When magnetic field diminishes
to 100~nG the GZK cutoff appears (dotted line), as it should according to
the propagation theorem.  

A weak GZK cutoff in the case of diffusive propagation is explained by
the flux suppression at $E<E_{\rm GZK}$ cutoff due to longer
propagation time from the source (see Section \ref{diff_diff} for the
detailed explanation).

The spectra calculated in this paper in the diffusion
approximation are compatible, for the relevant parameters, with the
numerical simulations in Yoshiguchi et al. (2003). As an example we
present in Fig. \ref{fig8} our spectra from a single source
calculated in the diffusion approximation for magnetic
configuration ($B_0,~l_c$)=(100~nG,~1~Mpc) with the Kolmogorov
spectrum of turbulence for a distance to the source of 10~Mpc (left
panel) and 30~Mpc (right panel). For the distance 10~Mpc we also plot
the spectrum for the rectilinear propagation shown by the dotted
curve. The two curves intersect at $E_* \sim 3\times 10^{20}$~eV,
as they should provided by the condition $l_d(E)=r$, where $r$ is
the distance to the source. At $E>E_*$ the particles propagate
rectilinearly  and the flux is given by the dashed line. This spectrum
is compared with the numerical simulations of Yoshiguchi et al. (2003)
for the same magnetic field configuration (Yoshiguchi (2004)),
shown by the crosses. Since the calculations of Yoshiguchi et al. (2003)
are not normalized, we equate the fluxes at energies with
rectilinear propagation as shown in Fig. \ref{fig8}. One can
observe considerable disagreement at $E \sim 8\times 10^{19}$~eV,
where it is possible to  suspect a transitional regime between
quasi-rectilinear and diffusive propagation in numerical
simulations. For the distance to the source r=30~Mpc (right panel)
the agreement is much better, probably because at these distances
the diffusion regime is reached in the numerical simulations.

The study of this paper is not intended to be a realistic one. We
consider the diffusive propagation of UHE particles using an analytic
approach with the aim of understanding the basic properties of
propagation in strong magnetic fields. A realistic propagation
should be studied within the hierarchical model of different
magnetic fields in different large scale structures, as it is
done in simulations (Sigl et al. (2003), Sigl et al. (2004), Dolag
et al. (2003)). However, we are able to reach some practical
conclusions.

Our study is focused on the  spectra of UHECR. We demonstrated
that the crucial parameter is the source separation $d$. For
a wide class of magnetic field configurations, when the separation
$d$ is smaller than the diffusion length $l_d$, the spectrum is
universal, i.e. the same as in case of rectilinear propagation.
For the simulation of Dolag et al. (2003) the spectrum must be
universal. The simulations of Sigl et al. (2003), Sigl et al.
(2004) and Yoshiguchi et al. (2003) include diffusive and
intermediate regimes. 
According to our calculation, in the diffusion regime very strong 
magnetic fields, $B_0 \sim 1000$~nG,  and large
separation between sources are needed to produce a spectrum with
weak (or absent) GZK cutoff. In this case very high source
luminosities are required, $L_p > 1\times 10^{47}$~erg/s. 
It disfavors   
diffusion in strong magnetic fields as the  explanation for the AGASA
excess at high energies.


\section*{Acknowledgments}
We acknowledge participation of Askhat Gazizov at an early stage of
this work. We are grateful to Vladimir Ptuskin for valuable
discussion of diffusive propagation of UHECR and to Igor Tkachev
for early discussion of hierarchical magnetic field structure of
the universe and the possibility of quasi-rectilinear propagation of
UHE protons. We acknowledge useful discussions with Pasquale Blasi 
and Yurii Eroshenko. We are grateful to the authors of the work Yoshiguchi et
al. (2003), especially Hiroyuki Yoshiguchi and Katsuhiko Sato, for
discussions and calculations made for comparison with our results.
We also thank Richard Ford for a critical reading of the manuscript.\\
We thank the transnational access to research infrastructures 
(TARI) program trough the LNGS TARI grant contract HPRI-CT-2001-00149.

\appendix\section{Spectral energy density and scale dependence of
magnetic fields.}

We shall derive here $\langle B^2 \rangle \propto k^{-s}$ from the
spectral energy density of a turbulent plasma $w(k) \propto k^{-m}$.
The turbulence is assumed to be described as an ensemble of
hydromagnetic waves with wave numbers $\vec{k}$ and with vanishing
mean electric field. The mean magnetic field on each scale
$\lambda=2\pi/k$ is $\langle \vec{B}\rangle_{\lambda} =
\vec{B}_{\lambda}$. i.e. on the scale $\lambda$ the magnetic field is
locally homogeneous. For the Alven waves the kinetic energy of
turbulent fluid is equal to magnetic energy (Landau and Lifshitz
(1987)).

Let us write down the Fourier expansion for the wave magnetic field. In the
limit $u_A/c \ll 1$, where $u_A$ is the Alven velocity, $\omega/ku_A\ll 1$
and the magnetic fields can be considered as quasi-static. Here and below
we shall omit the coefficients $(2\pi)^3$, which are inessential for our
discussion.
$$
\vec{B}(\vec{r},t)=\int d^3k \vec{B}(\vec{k},t)e^{i\vec{k}\vec{r}}
$$
\begin{equation}
\vec{B}^*(\vec{r},t)=\int d^3k' \vec{B}^*(\vec{k}',t)e^{-i\vec{k}'\vec{r}}
\label{expansion}
\end{equation}
The energy of magnetic field in the normalizing volume $V$ is
\begin{equation}
W=\frac{1}{8\pi}\int d^3r\vec{B}(\vec{r},t)\vec{B}^*(\vec{r},t)
\label{energy}
\end{equation}
Putting Eqs.(\ref{expansion}) into Eq.(\ref{energy}) and using
the definition of $\delta$ function
\begin{equation}
\int d^3r e^{i(\vec{k}-\vec{k}')\vec{r}}=\delta^3(\vec{k}-\vec{k'}),
\label{delta}
\end{equation}
we obtain
\begin{equation}
W=\frac{1}{8\pi}\int d^3k|B(\vec{k},t)|^2.
\label{W}
\end{equation}
Assuming $B({\vec{k},t})=B(k,t)$ we find the energy density of waves
and fluid  as
\begin{equation}
w =\frac{1}{2V}\int dk~ k^2 B^2(k).
\label{omega}
\end{equation}
The spectral energy density is then
\begin{equation}
w(k) =\frac{1}{2V} k^2 B^2(k).
\label{w_spectr}
\end{equation}
and hence for the Fourier component $B^2(k)\propto k^{-s_F}$,~~ $s_F=m+2$. 
Note that the
Fourier component $B(k)$ does not have meaning of the magnetic field strength 
over the scale $k$; moreover the dimension of $B(k)$ is different from a
magnetic field. The connection between $B_{\lambda}$, the average field on 
the scale $\lambda$, and the Fourier component $B(k)$, can be readily found
from relation
$$
w=\frac{1}{8\pi}B_{\lambda}^2=\frac{1}{2V}\int_{\lambda} dk~k^2B^2(k)
$$
where the integral is to be evaluated over region $\sim \lambda$. It results
in
\begin{equation}
B_{\lambda}^2 \sim \frac{4\pi}{V} k^3 B^2(k)=8\pi kw(k).
\end{equation}
Using the definition $B^2_{\lambda} \propto k^{-s_{\lambda}}$,  
we obtain for the physical field $B_{\lambda}$,~  $s_{\lambda}=m-1$,
and $s_F-s_{\lambda}=3$.

In particular, for the Kolmogorov spectrum we have $s_F=11/3$ and
$s_{\lambda}=2/3$, and thus for deflection of UHE particles
one should use the wave-number spectrum of magnetic fields in the form
\begin{equation}
B_{\lambda}^2 \propto k^{-2/3}~.
\label{spectrum}
\end{equation}


\begin{thebibliography}{00}

\bibitem{Arzimo}
L.A. Arzimovich and R.Z. Sagdeev, Physics of Plasma for Physicists
(in Russian), Moscow, Atomizdat, 1979.

\bibitem{Bere88}
V.S. Berezinsky and S.I. Grigorieva, A $\&$ A {\bf 199} (1988) 1.

\bibitem{Bere90a}
V.S. Berezinsky, V.A. Dogiel and S.I. Grigorieva, A$\&$A {\bf 232},
582 (1990a).

\bibitem{Bere90b}
V.S. Berezinsky, S.V. Bulanov, V.A. Dogiel, V.L. Ginzburg, and
V.S. Ptuskin, Astrophysics of Cosmic Rays, chapter IX, North-Holland (1990b).

\bibitem{Bere02a}
V. Berezinsky, A.Z. Gazizov and S.I. Grigorieva (2002a), hep-ph/0204357.

\bibitem{Bere02b}
V.S. Berezinsky, A.Z. Gazizov, and S.I. Grigorieva (2002b), astro-ph/0210095.

\bibitem{Bere03}
V. Berezinsky, A.Z. Gazizov and S.I. Grigorieva, hep-ph/0302483.
Proc. of Int. Workshop ``Extremely High Energy Cosmic Rays''
(eds M.Teshima and T.Ebisuzaki), Universal Academy Press, Tokyo 63, 2003.

\bibitem{Blasi99a}
P. Blasi, S. Burles, A.V. Olinto, Ap.J. {\bf 514}, 79 (1999a).

\bibitem{Blasi99b}
P. Blasi and A.V. Olinto Phys. Rev. {\bf D59} 023001 (1999b).

\bibitem{Carilli}
C.L. Carilli and G.B. Taylor, Annual Rev. Astr. Astroph., {\bf 40}, 319 (2002).

\bibitem{Casse}
F. Casse, M. Lemoine, G. Pelletier Phys. Rev. {\bf D65} 023002 (2002).

\bibitem{Dave}
R. Dav\`e, R. Cen, J.P. Ostriker, G.L. Bryan, L. Hernquist, N. Katz,
D.H. Weinberg, M.L. Norman and B. O'Shea, Astrophys.J. 547 (2001) 574.

\bibitem{Deli}
O. Deligny. A. Letessier-Selvon, E. Parizot, astro-ph/0303624.

\bibitem{Dolag}
K. Dolag, D. Grasso, V. Springel, I. Tkachev, astro-ph/0310902.

\bibitem{Giller}
M. Giller, J. Wdowczyk, A.W. Wolfendale, J.Phys. G {\bf 6} 1561 (1980).

\bibitem{Glushkov}
A.V. Glushkov and M.I. Pravdin, Astronomy Lett. {\bf 27}, 493 (2001).

\bibitem{Harari}
D. Harari, S. Mollerach, E. Roulet, JHEP 0207 006 (2002).

\bibitem{Hayashi96}
N. Hayashida et al. (AGASA collaboration), 
Phys. Rev. Lett. {\bf 77} 1000 (1996).

\bibitem{Hayashi99}
N. Hayashida et al. (AGASA collaboration), Ap.J., {\bf 522}, 225 (1999).

\bibitem{Isola}
C. Isola, M. Lemoine and G. Sigl, Phys. Rev. {\bf D65} 023004 (2002).

\bibitem{Kolmo}
A.N. Kolmogorov, Doklady Akad. Nauk. USSR, {\bf 30} 299 (1941).

\bibitem{Kraich}
R.N. Kraichnan, Phys. Fluids {\bf 8} 1385 (1965).

\bibitem{Kronberg}
P.P. Kronberg, Rep. Progr. Phys. {\bf 57}, 325 (1994).

\bibitem{Landau}
L.D. Landau and E.M. Lifshitz, Electrodynamics of Continuous Media,
chapter VIII, Magnetic Hydrodynamics, Pergamon Press, 1987.

\bibitem{Lemoine}
M. Lemoine, G. Sigl, P. Biermann, astro-ph/9903124.

\bibitem{Lifshi}
E.M. Lifshitz and L.P. Pitaevskii, Physical Kinetics
(v.10 of Theoretical Physics by L.D. Landau and E.M. Lifshitz)
Fizmatlit 2001.

\bibitem{Ryu}
D. Ryu, H. Kang, P. Biermann, Astron. Astroph., {\bf 335}, 19 (1998).

\bibitem{Sigl99}
G. Sigl, M. Lemoine, P. Biermann, Astrop. Phys. {\bf 10} 141 (1999).

\bibitem{Sigl03}
G. Sigl, F. Miniati and T.A. En{\ss}lin, Phys. Rev. {\bf D68} 043002 (2003).

\bibitem{Sigl04}
G. Sigl, F. Miniati and T.A. En{\ss}lin, astro-ph/0401084.

\bibitem{Spergel}
D.N. Spergel et al (WMAP collaboration), 
Astrophys. J. Suppl. {\bf 148} (2003) 175.

\bibitem{Stanev}
T. Stanev et al, Phys. Rev. {\bf D 62} 093005 (2000).

\bibitem{Syrov}
S.I. Syrovatskii, Sov. Astron. {\bf 3} 22 (1959).

\bibitem{Tinya}
P.G. Tinyakov and I.I. Tkachev, JETP Lett., {\bf 74} 445 (2001).

\bibitem{Uchiori}
Y. Uchiori et al, Asrop. Phys. {\bf 13}, 157 (2000).

\bibitem{Vainste}
S.I. Vainstein, A.M. Bykov and I.N. Toptygin, Turbulence, Stream
Layers and Shock Waves (in Russian), Nauka, Moscow (1989).

\bibitem{Vallee}
J.P. Vallee, Fund. Cosm. Phys. {\bf 19}, 1 (1997).

\bibitem{Wodow}
J. Wdowczyk, A.W. Wolfendale, Nature, {\bf 281}, 356 (1979).

\bibitem{Yoshi03}
H. Yoshiguchi, S. Nagataki, S. Tsubaki and  K. Sato, 
Astrophys.J. 586 (2003) 1211-1231.

\bibitem{Yoshi04}
H. Yoshiguchi, private communication.

\end{thebibliography}
\end{document}